\documentclass[aps,twocolumn,pra,superscriptaddress]{revtex4-2}

\usepackage{multirow}
\usepackage{booktabs}

\usepackage{soul}
\usepackage[T1]{fontenc}
\usepackage{dsfont}
\usepackage{amsfonts}
\usepackage{amsmath}
\usepackage{bm}
\usepackage{amssymb}
\usepackage{graphicx}
\usepackage{bm}%
\usepackage{braket}

\usepackage[colorlinks=true,urlcolor=blue,citecolor=blue,linkcolor=blue]{hyperref}
\DeclareGraphicsExtensions{.eps,.ps,.pdf}
\newcommand{\cte}{w}

\newcommand{\diff}{\mathop{}\!\mathrm{d}}
\newcommand{\al}{\alpha}

\newcommand{\la}{\lambda}
\newcommand{\ka}{\kappa}

\newcommand{\si}{\sigma}

\newcommand{\La}{\Lambda}

\newcommand{\BM}{\begin{displaymath}}

\newcommand{\EM}{\end{displaymath}}

\newcommand{\ie}{\hbox{\em i.e.{}}}

\def\nn{\nonumber}

\def \baa {\begin{eqnarray*}}
\def \eaa {\end{eqnarray*}}
\def \bb  {\begin{thebibliography}}
\def \eb  {\end{thebibliography}}

\def \lab #1 {\label{#1}}

\def \Tr {\text{Tr}}

\newcommand{\Rr}{\mathsf{R}}

\newcommand{\co}{\mathcal{C}}

\newcommand{\sjs}[6]{\left\{ \begin{array}{c c c}
#1 & #2 & #3
\\
#4 & #5 & #6
\end{array}
\right\} }

\newcommand{\da}{\dagger}

\newcommand{\con}[1]{\mathcal{C}_{#1} }

\newcommand{\rema}[1]{\textcolor{black}{#1}}
\newcommand{\remaRe}[1]{\textcolor{black}{#1}}

\makeatletter
\g@addto@macro\bfseries{\boldmath}
\makeatother

\usepackage[dvipsnames]{xcolor}

\newcommand{\gf}{\gamma}
\newcommand{\TRS}{\mathcal{T} }

\usepackage[normalem]{ulem}
\usepackage{xcolor}
\usepackage{soul}
\setstcolor{red}
\usepackage{enumitem}   

 \newcommand\redout{\bgroup\markoverwith
{\textcolor{red}{\rule[.5ex]{2pt}{2.4pt}}}\ULon}
 \newcommand\blueout{\bgroup\markoverwith
{\textcolor{blue}{\rule[.5ex]{2pt}{2.4pt}}}\ULon}
  
\begin{document}
\widetext

\title{ Thermal effects on the spin domain phases of high spin-$f$ Bose-Einstein condensates with rotational symmetries}

\author{Eduardo Serrano-Ens\'astiga}
\email{ed.ensastiga@uliege.be}
\affiliation{Departamento de F\'isica, Centro de Nanociencias y Nanotecnolog\'ia, Universidad Nacional Aut\'onoma de M\'exico\\
Apartado Postal 14, 22800, Ensenada, Baja California, M\'exico}
\affiliation{Institut de Physique Nucléaire, Atomique et de Spectroscopie, CESAM, University of Liège
\\
B-4000 Liège, Belgium}

\author{Francisco Mireles}
\email{fmireles@ens.cnyn.unam.mx}
\affiliation{Departamento de F\'isica, Centro de Nanociencias y Nanotecnolog\'ia, Universidad Nacional Aut\'onoma de M\'exico\\
Apartado Postal 14, 22800, Ensenada, Baja California, M\'exico}
\date{\today}
\begin{abstract}
Spinor Bose Einstein condensates (BEC) can be realized  nowadays using different atomic species of several spin values, offering unprecedented opportunities to scrutinize the underlying physics of its spin phase domains and of its quantum phase transitions. At sufficient low temperatures, 
lower than the critical temperature,  
a fraction of thermally excited atoms of the condensate can still interact with the whole system leading to spin-dependent interactions that can modify the nature of its phase domains. In this work, we characterize the thermal fraction of atoms of a spinorial BEC of general spin-$f$ value, provided that its ground state lies in a given spin phase with rotational symmetry. 
To that end, we use the Hartree-Fock approximation and \rema{a method} based on the Majorana stellar representation for mixed quantum states and symmetry arguments. We consider the spin phases with usual point group symmetries, including those with some exotic phases associated to the platonic solids. The method leads to useful analytical expressions of the eigenspectrum of the thermal cloud allowing us to study the admissible regions and multipolar magnetic moments of the spin phases as a function of the temperature for general spin values. 
\end{abstract}
\maketitle
\section{Introduction}
\label{sec.Int}
Interacting many-body systems with internal spin degree of freedom, such as unconventional superconductors, Helium-3 superfluids, and  ultracold atoms, constitute prototypical examples of physical systems where exotic and fascinating spin phases can be realized \cite{Vollhardt:1990,Kruchinin:2011,lewenstein2012ultracold,Kaw.Ued:12}.Owing the astonishing manipulation currently achievable in the laboratory, among the variety of spin systems in nature, the spinor Bose-Einstein condensates (BEC) of atomic cold gases confined through optical trapping, conforms an ideal platform to investigate such spin phases~\cite{Schmal.Erhard.etal:2004,Nay.Bre:16,PhysRevLett.119.050404,jimenez2019spontaneous}.  The first spinor BEC was realized experimentally in 1998 with $^{23}$Na atoms of $f=1$ \cite{stenger1998spin}, since then spinorial BECs has been already realized for spins $f=2,3,4,6$ and $8$~\cite{Chang.Hamley.etal:2004,PhysRevA.77.061601,Web.Her.Mar:03,lu2011strongly,PhysRevLett.108.210401,Schmal.Erhard.etal:2004}.

The spin phases of a BEC can vary not only with respect to the atomic species of the condensate, but also with applied external fields and temperature~\cite{Kaw.Ued:12,Kawa.Phuc.Blakie:2012,Pas.Mar.Ver:12,Nay.Bre:16}. At temperatures close to absolute zero, the atomic gas is very well described by a single macroscopic quantum state, description that remains valid even in the presence of external fields. However, as the temperature is increased the cloud of thermally excited atoms induces non-trivial spin-spin interactions \cite{griffin2009bose,Kawa.Phuc.Blakie:2012,Ser.Mir:21} leading to the appearance of a number of interesting phenomena, ranging from the appearance of magnetic spin domains, shifts in the phase boundaries, the arise of metastable phases, and to quantum quench dynamics, just to mention a few~\cite{jimenez2019spontaneous,PhysRevA.78.023632,sta.mie.chi:1999,sad.eto:2013,Ser.Mir:21,PRA.100.013622,PhysRevA.99.023606}. Theoretically, one of the simplest methods to study a many-body systems at finite temperatures as the BECs is the Hartree-Fock (HF) approximation~\cite{griffin2009bose,blaizot1986quantum}. The HF approach starts by assuming that the whole condensate at finite temperature can be partitioned in two parts described by a condensate fraction and a thermal cloud of noncondensate atoms. Both fractions are then determined via coupled Gross-Pittaevskii (GP) and HF equations which are usually solved numerically in a self-consistent manner~\cite{Kawa.Phuc.Blakie:2012}. However, the analytical derivations and numerical computations for the noncondensate fraction becomes quickly challenging as one increases the spin $f-$value since its total degrees of freedom increases as $(2f+1)^2$. An alternative way to overcome this problem is to take advantage of the following two results: {\it i}) most spin phases of BEC predicted by mean-field theory have symmetries in common with the Hamiltonian \cite{Bar.Tur.Dem:06,Mak.Suo:07,PhysRevA.84.053616,Kaw.Ued:12} and {\it ii}) the HF theory keep the symmetries \emph{self-consistent}~\cite{blaizot1986quantum}, \ie, the noncondensate fraction inherits the symmetries in common of the condensate fraction and the Hamiltonian. 
Hence, as we show {\it a posteriori} 
, we are able to reduce significantly the degrees of freedom describing the condensate, allowing us in turn, to gain further insight of its physical properties such as the magnetization and the atom populations at finite temperatures. In addition it allows us to inspect the admissible regions of the spin phases that leads to metastable phases and quench dynamics \cite{Ser.Mir:21,PRA.100.013622,PRL.124.043001}. 
 
In this work, we implement 
\rema{a method based in the Majorana representation~\cite{Maj.Rep}} to determine and fully characterize the noncondensate fraction of atoms for a general spinor BEC having a given  point group symmetry. In particular, we draw our attention to the family of spin phases of a general total spin $f=2,3,4$ and 6 that includes the Ferromagnetic, Polar, and Nematic phases, as well as the more exotic phases that are closely related to the platonic solids: Tetrahedron, Octahedron, Cube and Icosahedron. The latter phases are of wide interest owing its connection to the appearance of non-Abelian topological excitations~\cite{kasamatsu2005vortices,Koba.Kawa.Nitta.Ueda:2009,Kaw.Ued:12,lewenstein2012ultracold}. 
The paper is organized as follows: In Sec.~\ref{Sec.Theory} we review the Hamiltonian model used for a general spinor BEC, as well as the mean-field and the Hartree-Fock approximations employed to describe the system at zero and finite temperatures. We also discuss in this section the Majorana representation of pure and mixed spin states that allows to determine the spin phases with a given symmetry. The characterization of several noncondensate fractions of spinor BECs, and the physical implications on their multipolar magnetic moments, are explained in Sec.~\ref{Sec.Sphases}. The spin-2 BEC, which is the first spin value that includes a platonic phase, is studied with further detail in Sec.~\ref{Sec.Finite}, where especial attention is put on the admissible region of each spin-2 phase. We end the paper with some final comments and conclusions in Sec.~\ref{Sec.Conc}.
\section{Theory}
\label{Sec.Theory}
\subsection{Spinor BEC Hamiltonian}
We consider a BEC of an atomic gas with total spin $f$ confined through an optical trap. The condensate is assumed to be weakly interacting and sufficiently diluted such that only two-body collisions are predominant and the $s$-wave approximation is still valid~\footnote{\remaRe{The particle-particle interaction terms of the Hamiltonian should be given by a product of field operators in different positions  $ V_{p-p}(\bm{r_1} , \bm{r_2}) \psi_i^{\dagger} (\bm{r}_1) \psi_j^{\dagger} (\bm{r}_2) \psi_k (\bm{r}_1) \psi_l (\bm{r}_2)$. However, since we only consider point-contact interactions, $V_{p-p}(\bm{r_1} , \bm{r_2}) \propto \delta (\bm{r}_1 - \bm{r}_2)$. Then, the double integration over the $\bm{r}_1$ and $\bm{r}_2$ spatial variables is reduced to just one, as is written in Eq.~\eqref{Full.Ham}. See Ref.~\cite{Kaw.Ued:12,lewenstein2012ultracold} for more details}}. The spinor-quantum field associated to the spinor condensate is denoted by  $\hat{\bm{\Psi}}= (\hat{\psi}_f \, , \hat{\psi}_{f-1} \, , \dots \, , \hat{\psi}_{-f})^{\text T}$, where $\hat{\psi}_m$ are the field operators for a magnetic quantum number $m$, and T denotes the transpose. Then the full Hamiltonian in the second quantization formalism reads~\cite{Kaw.Ued:12,lewenstein2012ultracold}
\begin{align}
\hat{H} = & \hat{H}_{s} + \hat{V} 
= \int \diff \bm{r} \left\{ \sum_{i,j} \left[ \hat{\psi}_i^{\da} \left( h_s \right)_{ij} \hat{\psi}_j 
\right] + \hat{v} \right\} \, ,
\end{align}
where $h_s$ is the spatial contribution 
\begin{equation}
h_{s} = \left( - \frac{\hbar^2 \nabla^2}{2M}  + U(r) \right)\mathds{1}_{2f+1}
\, , \quad 
\end{equation}
with $M$ the atomic mass,  $U(r)$ is the potential energy associated to the optical trap and $\mathds{1}_{2f+1}$ is the $(2f+1)\times(2f+1)$ unit matrix. The interparticle interactions, included in $\hat{v}$, are also described in the second-quantization formalism and are usually written in terms of the interaction channels with coupling factors $c_{\gf}$ having $\gf=0, \, 1 , \dots f$~\cite{Kaw.Ued:12,lewenstein2012ultracold}.   Explicitly, the interaction terms are comprised by \begin{equation}
\label{Full.Ham}
\hat{v} 
= \sum_{\gf=0}^{f} \frac{c_{\gf}}{2} \mathcal{M}^{(\gf)}_{ijkl} \hat{\psi}_i^{\da} \hat{\psi}_j^{\da} \hat{\psi}_k \hat{\psi}_l  \, ,
\end{equation}
where $\mathcal{M}^{(\gf)}$ are numerical complex tensors independent of the atomic specie. For instance, in  the case of spin-2 BEC, there exists only three types of interacting terms \cite{Kaw.Ued:12}
\begin{align}
\label{v}
\hat{v} =&
\,  \frac{c_0}{2} : \hat{n}^2 :
+  \frac{c_1}{2} : \hat{\bm{F}}^2 :
+  \frac{c_2}{2} \hat{A}_{00}^{\da} \hat{A}_{00} \, ,
\nonumber
\\
=& \frac{c_0}{2} : \left( \sum_{i} \hat{\psi}_i^{\da} \hat{\psi}_i \right)^2 : + \frac{c_1}{2} : \left(  \sum_{\al,i,j} 
(F_{\al})_{ij} \hat{\psi}_i^{\da} \hat{\psi}_j \right)^2 :
\nonumber
\\
& + \frac{c_2}{10}\left( \sum_{i} (-1)^i \hat{\psi}^{\da}_i \hat{\psi}^{\da}_{-i} \right) \left( \sum_{i} (-1)^i \hat{\psi}_i \hat{\psi}_{-i} \right)
\nonumber
\\
=& 
\, \frac{c_0}{2} \sum_{i,j} \hat{\psi}_i^{\da} \hat{\psi}_j^{\da} \hat{\psi}_j \hat{\psi}_i
+ \frac{c_1}{2} \sum_{\al,i,j,k,l} 
(F_{\al})_{ij} (F_{\al})_{kl} \hat{\psi}_i^{\da} \hat{\psi}_k^{\da} \hat{\psi}_l \hat{\psi}_j
\nonumber
\\
& + \frac{c_2}{10}\left( \sum_{i} (-1)^i \hat{\psi}^{\da}_i \hat{\psi}^{\da}_{-i} \right) \left( \sum_{i} (-1)^i \hat{\psi}_i \hat{\psi}_{-i} \right)
\, ,
\end{align}
 where $: \, \,:$ denotes the normal ordering of the field operators and $F_{\al}$ the angular momentum matrices components of spin $f=2$ along the $\al=x,y$ or $z$ axes,  written in units of $\hbar$. Hence the entries of the tensors $\mathcal{M}^{(\gf)}$ are 
\begin{align}
\mathcal{M}^{(0)}_{ijkl} = & \delta_{il} \delta_{jk} \, , \quad 
\mathcal{M}^{(1)}_{ijkl} = (F_{\al})_{il} (F_{\al})_{jk}
\, ,
\nonumber
\\
& \mathcal{M}^{(2)}_{ijkl} = \frac{(-1)^{i+k}}{5}  \delta_{i,-j} \delta_{k,-l} 
\, .
\end{align}
On the other hand, the coefficients $c_{\gf}$ in (\ref{v}) are linear combinations of the $s$-wave scattering lengths of the total spin-$F$ channel $a_F$ $(F=0,2,4)$ whose values will depend on the atomic species of the condensate~\cite{Kaw.Ued:12,lewenstein2012ultracold}. In particular the $c_{\gf}$ coefficients for the spin-2 BEC case are given by \cite{Kaw.Ued:12}
\begin{equation}
\label{f2.coupling}
c_0 = \frac{4g_2 + 3g_4}{7} , \, \, c_1 = \frac{g_4 - g_2}{7} ,  \, \,
c_2 = \frac{7g_0 - 10g_2 + 3g_4}{7} ,
\end{equation}
with
\begin{equation}
g_F= \frac{4\pi \hbar^2}{M} a_F \, .
\end{equation}
We plot in Table \ref{tab.1} the $a_F$ values for spin-2 condensates of $^{23}$Na and two isotopes of Rb.
\begin{table*}[t!]
\begin{tabular}{p{2cm} p{2cm} p{2.5cm} p{2.5cm} p{2.5cm} p{2cm} p{2cm}}
\hline
 &  & \multicolumn{3}{c}{ Scattering length($a_B$)} &  \multicolumn{2}{c}{ Spin-dependent coupling factors }
\\
Atom & $M$(u) & $a_4$ & $a_2$ & $a_0$ & $c_1/c_0$ & $c_2/c_0$
\\
\hline
$^{23}$Na & $22.99 $  & $64.5 \pm 1.3$ & $45.8 \pm 1.1$ & $34.9\pm 1.0$   & $0.05 \pm 0.005$ & $-0.05 \pm 0.04$ 
\\
$^{87}$Rb &  $86.91$  & $106.0 \pm 4.0$ &  $94.5 \pm 3.0$ & $89.4 \pm 3.0$ & $0.017\pm0.007$ & $-0.002 \pm 0.06^{\text{a}}$
\\
$^{83}$Rb &  $82.92$  & $81.0\pm3$ & $82.0\pm 3$ & $83.0 \pm 3$ & $-0.002\pm 0.007$ & $0.0070\pm 0.07^{\text{a}}$
\\
\hline
\end{tabular}
\caption{\label{tab.1} Scattering lengths $a_F$, extracted from \cite{ciobanu2000phase}, and ratios of the coupling factors $c_{\gf}$ \eqref{f2.coupling} of the atomic species  $^{23}$Na and two isotopes of Rb. $^{\text{a}}$Here, we decided to leave an extra decimal beyond to the uncertainty scale.
}
\end{table*}
 The term associated to $c_0$ is usually called spin-independent interaction since it is equivalent to the square of the number operator. Clearly, the rest of the interactions are spin-dependent. The Hamiltonian \eqref{Full.Ham} has a symmetry group isomorphic to $SO(3) \times \mathds{Z}_2 $ constituted by the group of rotations and the time-reversal symmetry.
\subsection{Mean-field approximation and Majorana representation for pure spin states}
In mean-field (MF) theory one considers that all the atoms in the spinor condensate are in the same quantum state described by a spinor order-parameter $\langle \hat{\bm{\Psi}} \rangle = \bm{\Phi}$ \cite{Kaw.Ued:12,lewenstein2012ultracold}. We assume for simplicity that the spatial and spinorial sectors of the system are factorizable \remaRe{and, consequently, they are mutually independent}. Besides, we consider \remaRe{the box potential with size $L$~\cite{PhysRevA.71.041604,pethick2008bose,lewenstein2012ultracold},} $U(\bm{r})=0$, where the ground state is $\phi_j (\bm{r}) = \phi_j e^{i  \bm{k}\cdot \bm{r}}$ with $\bm{k} = \bm{0}$ and the spinor order-parameter $\bm{\Phi}= (\phi_f ,  \phi_{f-1} , \dots , \phi_{-f})^{\text{T}}$. \remaRe{We consider $L$ tends to infinity such that the index $\bm{k}$ have a 3-dimensional domain $\bm{k} \in \mathds{R}^3$. We will use this result in Sec.~\ref{Sec.Finite}}. The condensate is thus constrained to a fixed \remaRe{density of particles} $N= \bm{\Phi}^{\da} \bm{\Phi}$ where the ground state $\bm{\Phi}$ of the BEC minimizes the functional MF energy $ E[\bm{\Phi}]= \langle \hat{H} \rangle$. It is known that the majority of the ground states of the BEC, without external fields, have rotational symmetries \cite{Bar.Tur.Dem:06,Mak.Suo:07,PhysRevA.75.023625,Kaw.Ued:12,PhysRevA.84.053616}. This result was predicted by Michel's theorem \cite{RevModPhys.52.617} which dictates that, for a real function $\mathcal{F}$ with domain $D$, $\mathcal{F}: D \rightarrow \mathds{R}$, the points $p \in D$ with symmetries in common with $\mathcal{F}$ may be critical. Consequently, one can find ground states of any spinor BEC by making use of the symmetry subgroups of rotations, as proposed in Ref.~\cite{PhysRevA.84.053616}. The symmetry point group of a given order parameter $\bm{\Phi}$ can be found through the Majorana representation for pure states \cite{Maj.Rep}, which is defined via a polynomial that involves the coefficients of  $\bm{\Phi}$, that reads explicitly
\begin{equation}
\label{first.pol}
p_{\psi}(Z) = \sum_{m=-f}^f (-1)^{f-m}   \sqrt{\binom{2f}{f-m}} \phi_m z^{f+m}  \, .
\end{equation}
The polynomial $p_{\psi}(z)$ has degree at most $2f$, and by rule, its set of roots $\{ \zeta_k \}_{k}$ is always increased to $2f$ by adding the sufficient number of roots at infinity. Now, we associate to each root $\zeta= \tan(\theta/2) e^{i \phi}$, via the stereographic projection from the south pole, a point on the sphere $S^2$ with spherical angles $(\theta \, , \phi)$. Hence, the \emph{constellation} $\co_{\psi}$ of $\ket{\psi}$ is thus defined as the set of $2f$ points on $S^2$, called \emph{stars}. 
 
When $\bm{\Phi}$ is transformed in the Hilbert space by the unitary representation  $D(\Rr )$ of a rotation $\Rr \in SO(3)$, the constellation rotates by $\Rr$ on the physical space $ \mathds{R}^3$. In addition, the time-reversal operator transforms the Majorana constellation of the spinor $\bm{\Phi}$ to the antipodal counterpart~\cite{beng17geo}. Therefore, the quantum state $\bm{\Phi}$ share the same point group symmetry of the polyhedron associated to the constellation $\con{\bm{\Phi}}$. More details about the Majorana representation and their applications are found in Refs.~\cite{beng17geo,Chr.Guz.Ser:18}.
 As an illustration, let us consider the spin-2 BEC and calculate its phase diagram in the parameter space of the spin-dependent coupling factors $(c_1, c_2)$. The application of MF theory in Eq.~\eqref{v} leads to the energy of the system 
 \begin{align}
E[\bm{\Phi}] = & \, \frac{c_0}{2} \left(\bm{\Phi}^{\da} \bm{\Phi} \right)^2
+ \frac{c_1}{2} \sum_{\al} \left( \bm{\Phi}^{\da} F_{\al} \bm{\Phi} \right)^2 
\nonumber
\\
&
+ \frac{c_2}{10} |\bm{\Phi}^{\dagger} \TRS \bm{\Phi} |^2
- \mu \left(  \bm{\Phi}^{\da} \bm{\Phi} -N \right) \, ,
 \label{energy.GS}
\end{align}
where we constrained the system to a fixed number of particles $N$ with Lagrange multiplier being the chemical potential $\mu$. The time-reversal operator $\TRS$ transforms the spinor $\bm{\Phi}$ as
\begin{equation}
\TRS \Phi_k = (-1)^{f+k} \Phi^*_{-k} \, . 
\end{equation}
For the following sections, we find convenient to write the  expression (\ref{energy.GS}) in terms of the density matrix \remaRe{of the condensate gas in the MF solution}, $\rho= \bm{\Phi} \bm{\Phi}^{\da}$ where $\Tr \, \rho =N$. Hence, 
\begin{align}
E[\bm{\Phi}] = & \, \frac{c_0}{2} \Tr \left[ \rho \right]^2 
+ \frac{c_1}{2} \sum_{\al} \Tr \left[ \rho F_{\al} \right]^2 
\nonumber
\\
& + \frac{c_2}{10} \Tr \left[ \TRS \rho \TRS \rho \right]
- \mu \left(  \Tr[\rho] -N \right) \, .
\end{align}
\begin{figure}[t]
\includegraphics[scale=0.65]{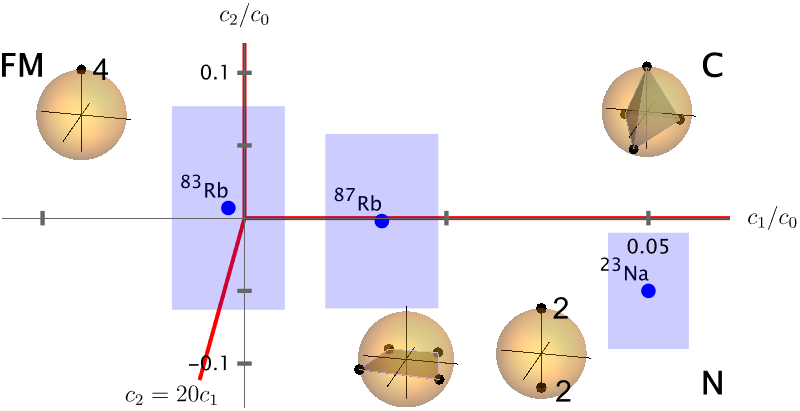}
\caption{
\label{Diag.pha} Diagram phase for spin-2 BECs obtained with MF theory. The Majorana representation of some states of each phase are shown. The adjacent number in some points correspond to its degeneracy. The values and their respective uncertainties of the atomic species of spin-2 (see Table~\ref{tab.1}) are marked on the figure. 
}
\end{figure}

\begin{figure}[t]
\begin{center}
\includegraphics[scale=0.4]{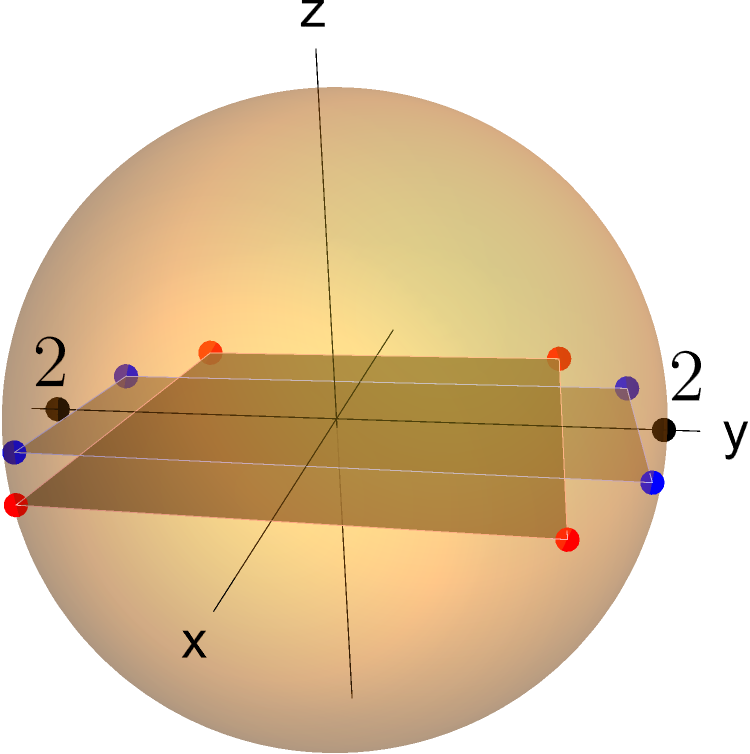}
\end{center}
\caption{
\label{Nem.pha} Majorana constellations of the nematic states of spin-2 BEC with $\eta= \pi/3 \, , 3\pi/8 \, , \pi/2$, shown in color black, blue and red, respectively.
}
\end{figure}

Note that the spin-independent interaction 
$\Tr \left[ \rho \right]^2$ contributes just as a constant term. On the other hand, the other two interactions are spin dependent and, consequently, able to modify the ground state of the system. The phase diagram of the spin-2 BEC in the parameter space $(c_1 , \, c_2)$ is plotted in Fig.~\ref{Diag.pha}. We shall now describe the four different ground states allowed:
\begin{itemize}
\item[1.] {\it Ferromagnetic} (FM) {\it phase}:  The spinor order-parameter has only one non-zero coefficient, $\phi_2 = \sqrt{N}$. It is symmetric under rotations about the $z$ axis, with point group isomorphic to the special orthogonal group $SO(2)$. Its constellation $\con{\bm{\Phi}}$ consists of $4$ coincident points on the north pole. Each atom is fully magnetized along the $z$ axes $M_z \equiv \langle F_z \rangle/N = 2$ and $M_x = M_y = 0$. Hence, the FM phase maximizes the $c_1$-interaction term $\langle \bm{F} \rangle^2=4N$, whereas its $c_2$-interaction vanishes $|\langle \TRS \rangle |^2=0$. The FM phase appears for any spin-$f$ condensate and corresponds to the $\ket{f,f}$ phase of the $\ket{f,m}$ family mentioned below.
\item[2.] {\it Nematic family}: It consists of a family of quantum states $\bm{\Phi}= \sqrt{N}(\sin \eta/ \sqrt{2} , 0 , \cos \eta , 0 , \sin \eta/ \sqrt{2} )^{\text{T}}$ parametrized by $\eta \in [\pi/3 \, ,\pi/2]$~\cite{Bar.Tur.Dem:06} (see Fig.~\ref{Nem.pha}). Contrary to the FM phase, this state minimizes the $c_1$-interaction, whiles maximizing the $c_2$-interaction, $\langle \bm{F} \rangle^2=0$ and $|\langle \TRS \rangle |^2=N$, respectively. \rema{There are two exceptional phases of the nematic family}:
\item[2a.] {\it Polar (P) phase}: Here $\eta=\pi/2$, yielding that $\phi_0 = \sqrt{N}$ and the other terms are equal to zero. Its symmetry group, \remaRe{denoted by $D_{\infty}$, consists of the group generated by any rotation about the $z$ axis, and a rotation by $\pi$ about any axis on the equator}. The constellation of the P phase has 2 points on each pole of the sphere. This phase exists for any spin-$f$ BEC, being the state $m=0$ of the $\ket{f,m}$ phases (See Sec.~\ref{Sec.Sphases}).
\item[2b.] {\it Square} (S) {\it phase}: It is obtained when $\eta=0$, and its non-zero order-parameter terms are $\phi_{2}= \phi_{-2} = \sqrt{N/2}$. Its Majorana constellation consists of a square. Hence, $\bm{\Phi}$ has the dihedral point group denoted by $D_4$ \cite{book.bra.cra:10}. This phase belongs to the class of NOON spin states that we describe in Sec.~\ref{Sec.Sphases}.
\item[3.] {\it Tetrahedron} (T) {\it phase}: This spin-2 phase has $\bm{\Phi}=\sqrt{N/3}(1,0,0,\sqrt{2},0)^{\text{T}}$.  The order-parameter has a constellation forming a tetrahedron. Hence, its symmetry group is the tetrahedron point group denoted by $T$ in the Schönflies notation \cite{book.bra.cra:10}. It has null magnetization and it is orthogonal to its antipodal state, yielding both spin-dependent interactions equal to zero, $\langle \bm{F} \rangle^2=|\langle \TRS \rangle |^2=0$. This phase is also called the cyclic phase C \cite{ciobanu2000phase,PhysRevA.65.063602}.
\end{itemize}
\remaRe{We remark that a generic nematic phase is a quantum superposition of the polar and square phases. An analogous degenerate family was found earlier by Mermin in the context of $d$-wave pairing~\cite{Mermin74}. }

The phases mentioned above also appear when we consider other terms in the Hamiltonian such as linear and quadratic Zeeman terms \cite{Kaw.Ued:12,ciobanu2000phase}. We plot in Fig.~\ref{Diag.pha} the Majorana constellation associated to each phase discussed above, including the square and polar constellations of the nematic family. Given the symmetry and magnetization properties mentioned, it is easy to produce the phase diagram of the ground state in the parameter space $(c_1,c_2)$ as shown in Fig. \ref{Diag.pha}. We analyze its nature in each region: 
%
%
%
%
%
\begin{enumerate}[label=(\roman*)]
\item Quadrant $c_1>0$ and $c_2>0$: The ground state is given by the solution that minimizes both interactions. In this region, the spinor order-parameter is the cyclic phase. 

\item Quadrant $c_1>0$ and $c_2<0$: The ground state is such that minimizes $\langle \bm{F} \rangle^2$ and maximizes $|\langle \TRS \rangle |^2$. The state that fulfills these conditions is the N phase.

\item Quadrant $c_1<0$ and $c_2>0$: In this case we have the opposite conditions that the ones mentioned in (2), which are fulfilled by the FM phase.

\item Quadrant $c_1<0$ and $c_2<0$: Here, we need to compare between the minimal values attained by both interaction terms, taking into account that $|\bm{\Phi}|^2=N$. The minimums of the $c_1$ and $c_2$ terms are $2c_1N^2$ and $c_2N^2/10$, respectively. Hence, the ground state is the FM phase when $20c_1 \leq c_2$ and the N phase otherwise.
\end{enumerate}
All the phase transitions are of first order since at least one of the order-parameter coefficients changes drastically over each phase boundary. 

\subsection{Hartree-Fock approximation and the Majorana representation for mixed states}
In BECs, the simplest many-body theory that goes beyond (MF) approximation, and is capable of capturing the relevant physics occurring in its spin phase diagrams, as well as in its concomitant boundary regions, is the Hartree-Fock (HF) approximation~\cite{blaizot1986quantum,griffin2009bose,Kawa.Phuc.Blakie:2012,Ser.Mir:21}.
Also, HF is robust when describing spinor condensates at finite, but ultra-cold temperatures~\cite{griffin2009bose,Kawa.Phuc.Blakie:2012,phuc2013beliaev}. Formally, in the HF approximation the field operator is given by the order parameter plus a variational perturbation, $\hat{\psi}_j = \phi_j + \hat{\delta}_j$. The physical picture is that the cold atomic gas is assumed to be formed by two portions: the condensate (c) and the noncondensate (nc) atomic fractions. Each fraction \remaRe{of the atomic gas} is then represented by a density matrix $\rho_{ij}^c = \phi_i \phi_j^* $ and $\rho_{ij}^c = \langle \hat{\delta}_j^{\dagger} \hat{\delta}_i \rangle$, composed of pure and a mixed quantum states, respectively. The atomic fraction of each part is given by $N^a = \Tr (\rho^a)$ for $a=n , \, nc$, and satisfy $N^c + N^{nc} = N$. The noncondensate atoms $\rho^{nc}$ act as a cloud of excited atoms that interacts non-trivially with the condensate fraction $\rho^c$. The total system is then denoted by $\rho =\rho^{c}+\rho^{nc}$ with $\Tr \rho = N$. A reasonable approximation of the approach is to neglect the anomalous density $\langle \hat{\delta}_i \hat{\delta}_j \rangle$ (Popov approximation), and the three-field correlations $\langle  \hat{\delta}_i \hat{\delta}_j \hat{\delta}_k^{\da} \rangle$, which is still valid for diluted atomic gases at finite ultracold temperatures below the condensation temperature~\cite{Griffin:96,Ser.Mir:21}. \remaRe{The maximum temperature where the HF theory predicts a spin phase can be estimated by the condensation temperature $T_c^{\text{spin}}$ of an ideal spin-$f$ BEC trapped in a box. It is known that $T_c^{\text{spin}}$ is equal to the condensation temperature of an ideal scalar gas $T_0$ rescaled by the internal spin-states
$T_c^{\text{spin}} = (2f+1)^{-2/3} T_0 $~\cite{Kawa.Phuc.Blakie:2012}, where $T_0 = 3.31 \hbar^2 N^{2/3}/(k_BM) $~\cite{pethick2008bose}, with $k_B$ the Boltzmann constant. As an example, for BEC with Na$^{23}$ atoms with a typical density achieved in experiments $N= 10^{14} cm^{-3}$~\cite{stamper.Andrews.etal:1998}, $T_0= 1.5 \mu$K, and then $T_c^{\text{spin}} \approx 0.72\mu$K , 0.51 $\mu$K for $f=1$ and 2, respectively. }

As we emphasize in the following sections, even in the scenario in which the fraction associated to $\rho^{nc}$ is small compared to that associated to $\rho^c$, new interactions of the type $\rho^{c}\leftrightarrow\rho^{nc}$ and $\rho^{nc}\leftrightarrow\rho^{nc}$ emerge owing the non-trivial collisions among the atoms. Consequently, important modification of the physical properties of the BEC can take place, such that strength changes of the magnetization, shifting of the phase-transition boundaries, and variations of the allowed region for the metastables phases~\cite{Kawa.Phuc.Blakie:2012,Ser.Mir:21}. Operationally, those new extra two-body interaction terms, called direct and exchange interactions, are calculated through the use of the Wick theorem (see e.g. p.~89 of Ref.~\cite{blaizot1986quantum})
\begin{align}
\langle \hat{\psi}_i^{\dagger} \hat{\psi}_j^{\dagger} \hat{\psi}_k \hat{\psi}_l \rangle
=& \, 
\rho_{ik}^c \rho_{jl}^c  + \rho_{ik}^{nc} \rho_{jl}^{nc} + \rho_{il}^{nc} \rho_{jk}^{nc}
\nonumber 
\\
& + \rho_{ik}^c \rho_{jl}^{nc} + \rho_{il}^c \rho_{jk}^{nc} + \rho_{jk}^c \rho_{il}^{nc} + \rho_{jl}^c \rho_{ik}^{nc} \, .  
\end{align}
Note that $\rho_{ik}^c \rho_{jl}^c= \rho_{jk}^c \rho_{il}^c$ since the condensate fraction is represented by a pure quantum state. As an example to the previous equation, we express the $c_{\gf}$ spin-interactions obtained in the HF theory for $\gf=0,1,2$~\eqref{v} which reads,
\begin{align}
 \langle \hat{\psi}_i^{\dagger} 
\hat{\psi}_j^{\dagger} \hat{\psi}_j \hat{\psi}_i \rangle \approx & \left\{ \Tr \left[\rho^c + \rho^{nc} \right] \right\}^2 + \Tr \left[ \rho^{nc} \left( 2\rho^c + \rho^{nc} \right) \right] 
\nonumber
\\
= & \, N^2 + \Tr \left[ \rho^{nc} \left( 2\rho^c + \rho^{nc} \right) \right]  \, ,
\label{c0.HF}
\end{align}
\begin{align}
& \sum_{\alpha} (F_{\alpha})_{il} (F_{\alpha})_{jk} \langle \hat{\psi}_i^{\dagger} \hat{\psi}_j^{\dagger} \hat{\psi}_k \hat{\psi}_l \rangle \approx
\nonumber
\\
& \sum_{\alpha} \left\{ \Tr \left[\rho F_{\alpha} \right]^2 
+ \Tr \left[ F_{\al} \rho^{nc} F_{\al} (2 \rho^c + \rho^{nc}) \right]
\right\} \, ,
\label{c1.HF}
\end{align}
\begin{align}
\langle \hat{A}_{00}^{\dagger} \hat{A}_{00} \rangle \approx
\frac{1}{2f+1} \Tr \left[ \TRS \rho \TRS \rho + \TRS \rho^{nc} \TRS (2\rho^c + \rho^{nc}) \right] \, ,
\label{c2.HF}
\end{align}
where we use the Einstein summation convention for the latin repeated indices. The time-reversal operator $\TRS$ acts in $\rho$ as
\begin{equation}
(\TRS\rho \TRS)_{ji} = (-1)^{2f-i-j} \rho_{-i-j} \, .
\end{equation}
 A generic  noncondensate fraction of a BEC of spin $f$ has $(2f+1)^2$ degrees of freedom. However, for systems with a self-consistent symmetry, \ie, for cases such that the nonperturbed phase $\rho^c$ and the Hamiltonian of the system $\hat{H}$ have some symmetries in common, the noncondensate fraction $\rho^{nc}$ inherits such symmetries, and when solving the GP equations of the BEC, its degrees of freedom are reduced significantly (see 
 \cite{blaizot1986quantum} for further details). This leaves us now to determine the mixed states $\rho^{nc}$ that share the common symmetries with the order-parameter $\bm{\Phi}$ and the full Hamiltonian \eqref{Full.Ham}. 
 To that aim, a generalization of the Majorana representation for mixed states \cite{Ser.Bra:20} and exploit the point group symmetries of the physical system.
 
Let $\mathcal{B}_f$ be the set of the density matrices of spin-$f$ states. Thus $\mathcal{B}_f$ is equivalent to the set of $(2f+1)\times (2f+1)$ complex matrices which satisfy the following properties \cite{beng17geo}:
\begin{enumerate}[label=(\alph*)]
\item  $\quad \rho^{\dagger} = \rho$\,, \,\,\,\,\,\, \quad (Hermiticity)  
\item  $\quad \Tr(\rho) = 1 \, , \quad$ (Unit trace)
\item  $\quad \rho \geq 0\, ,  \quad$ \,\,\,\,\,\,\,\,\,\,\,(Semi-positive condition).
\end{enumerate}
Observe that for a BEC, the total $\rho$, the condensate $\rho^c$ and the non condensate $\rho^{nc}$ fractions of the atoms belong to the set of mixed states with the exception that their traces are different, where $\Tr (\rho^a) = N^{a}$, where $a=n$, $nc$, or omitted for the whole system. 

In $\mathcal{B}_f$ exists an orthonormal basis given by the set of the tensor operators $T_{\sigma \mu}$ with $\si=0,\dots 2f$ and $\mu = -\si , \dots , \si$~\cite{Fan:53,BrinkSatchler68,Var.Mos.Khe:88}, which written in terms of the Clebsch-Gordan
coefficients $C_{j_1 m_1 j_2 m_2}^{j m}$, reads
\begin{equation}
\label{decomp.TensOp}
T_{\sigma \mu}
=
\sum_{m,m'=-f}^f (-1)^{f-m'} C_{fm,f-m'}^{\si \mu}\ket{f,m}\bra{f,m'} 
\, ,
\end{equation}
satisfying the following properties,
\begin{equation}
\Tr ( T_{\sigma_1 \mu_1}^{\dagger} T_{\sigma_2 \mu_2} ) = \delta_{\sigma_1 \sigma_2} \delta_{\mu_1 \mu_2} \, , \quad T_{\sigma \mu}^{\dagger} = (-1)^{\mu} T_{{\sigma}{-\mu}} \, .
\label{hermi.tens.op}
\end{equation}
The key property of this basis is that they transform  block-diagonally under a unitary transformation $U(\Rr)$, that represents a rotation $\Rr \in SO(3)$, according to an \emph{irrep} of $SO(3)$
 $D^{(\sigma)}(\Rr)$, such that
\begin{equation}
\label{prop.sh}
U(\Rr) T_{\sigma \mu} U^{-1}(\Rr) = \sum\limits_{\mu'=-\sigma}^{\sigma} D_{ \mu' \mu}^{(\sigma)}(\Rr) T_{\sigma \mu'} \, ,
\end{equation}
where $D^{(\si)}_{ \mu' \mu} (\Rr) \equiv \braket{\si , \, \mu' |
    e^{-i \al F_z} e^{-i \beta F_y} e^{-i \gamma F_z} | \si , \, \mu}$ is the Wigner D-matrix \cite{Var.Mos.Khe:88} of a rotation $\Rr$ with Euler angles $(\al, \, \beta , \, \gamma)$, and
$\sigma=0,1,2,\ldots$ labels the irrep. 

The noncondensate fraction $\rho^{nc}$ of a spinor BEC can thus be written in terms of this basis
\begin{equation}
\rho^{nc} = N^{nc} \left( \frac{\mathds{1}_f}{2f+1}+ \sum_{\si=1}^{2f} \bm{\rho}_{\si} \cdot \bm{T}_{\si} \right) \, ,
\label{bloc.decomp}
\end{equation}
where $\bm{\rho}_{\si}= (\rho_{\si \si} , \dots ,\rho_{\si -\si}) \in \mathds{C}^{2\si+1}$ with
$\rho_{\si \mu} = \Tr( \rho \, T^{\da}_{\si \mu})$, $\bm{T}_{\si}=
(T_{\si \si} , \dots , T_{\si , -\si})$ is a vector of matrices, and
the dot product is the short for $\sum_{\mu=-\sigma}^\sigma\rho_{\si \mu}T_{\si \mu} $.
The properties of the density matrices and the tensor operators imply that each
$\bm{\rho}_{\si}$ vector can be associated to a constellation \emph{\`a la Majorana} \eqref{first.pol} consisting of $2\si$ points on $S^2$. Besides, the hermiticity condition of $\rho^{nc}$ and Eq.~\eqref{hermi.tens.op} implies that the constellation of every $\bm{\rho}_{\si}$ has antipodal symmetry~\cite{Ser.Bra:20}. As it is explained in Ref.~\cite{Ser.Bra:20}, the norm and complex phase factor of each $\bm{\rho}_{\si}$ are relevant to specify $\rho^{nc}$. The norm of $\bm{\rho_{\si}}$, denoted by $r_{\si}$, can be associated to the radius of the sphere. On the other hand, by the hermiticity of $\rho^{nc}$, there are only two options, both differing by a minus sign~\cite{Ser.Bra:20}. We present in Appendix~\ref{App.phase} a method, based on Ref.~\cite{Ser.Bra:20}, to associate the phase factor to a certain equivalence class of points in the constellation. For our purposes to determine the noncondensate fraction with a particular point group, it is sufficient to add the choice of sign to the radius $r_{\si}$, despite that the sign is not considered to define the radius of the spheres. Therefore, a mixed state is associated to a set of $2f$ constellations, denoted by $\con{\rho^{nc}}$, with antipodal symmetry over spheres with radii $r_{\si}$, respectively. In particular, $\rho^{nc}$ has a point group $G$ if and only if all the constellations of the $\bm{\rho}_{\si}$ vectors share the same symmetry:
\begin{equation}
D^{(\si)}(g) \bm{\rho}_{\si} = \bm{\rho}_{\si} \, , \quad \text{for each } \, g\in G \, .
\end{equation}
%

\begin{figure}[t!]
\begin{tabular}{|c|c|}
\hline
$\mathcal{C}_{\bm{\Phi}} $ &
$\mathcal{C}_{\rho^{nc}}$
\\
\hline
\begin{tabular}{c}
$\ket{f,m}$
\\[0.2cm]
\includegraphics[scale=0.13]{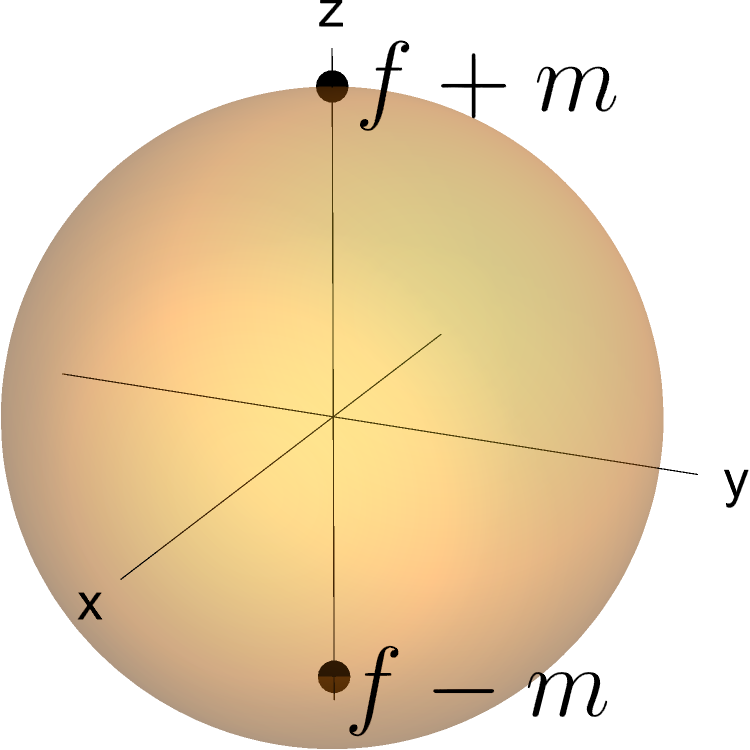}
\end{tabular}
&
\begin{tabular}{c}
\phantom{$\ket{f,m}$}
\\[0.2cm]
\includegraphics[scale=0.28]{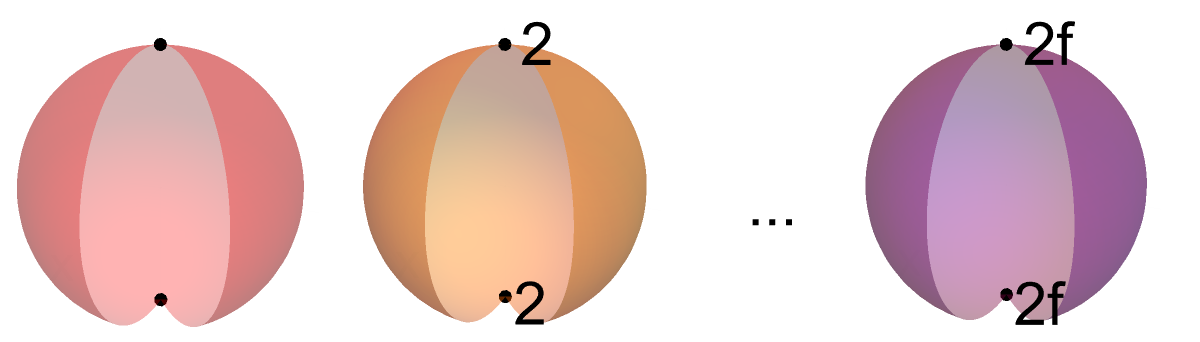}
\end{tabular}
\\
\hline
\begin{tabular}{c}
NOON spin
\\
\includegraphics[scale=0.16]{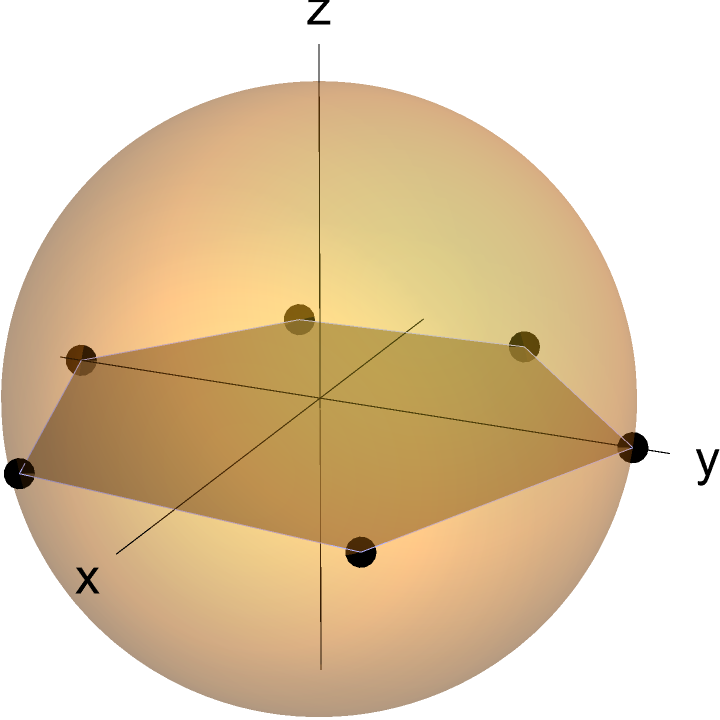}
\\
{\scriptsize $2f$-sided}
\\[-0.1cm]
{\scriptsize polygon}
\end{tabular} 
& 
\begin{tabular}{c}
\phantom{NOON spin}
\\
\includegraphics[scale=0.28]{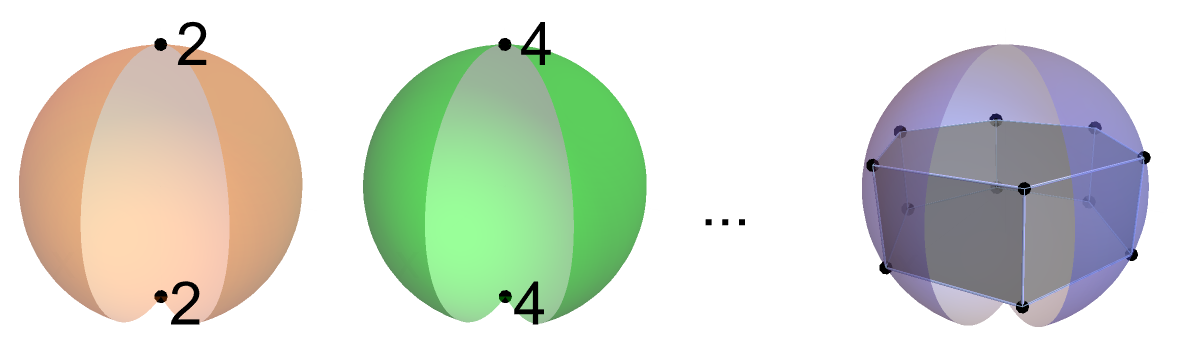}
\\
\hspace{4.2cm} {\scriptsize $2f$-faced}
\\[-0.1cm]
\hspace{4.2cm} {\scriptsize prism}
\end{tabular}
\\
\hline
\end{tabular}
\caption{\label{MSC.FFNOON} Majorana representation of mixed states of the families $\ket{f,m}$ and NOON spin states.
} 
\end{figure}
\section{Spin phases and magnetic moments}
\label{Sec.Sphases}
Let us now analize the set of mixed states with a specific point group symmetries. We study three families of spin phases: A) $\ket{f,m}$ states, B) NOON states, and C) the spin phases associated with a platonic solid. Their respective Majorana constellations of $\rho^{nc}$ are summarized in Figures~\ref{MSC.FFNOON} and \ref{MSC.Platonic}. We also calculate its corresponding eigenspectrum for the cases that it is possible to obtain analytical expressions, and that will be helpful in the next section. In addition, we discuss the multipolar magnetic moments of the spin phases which are closely related, as we explain below, to the Majorana representations of each state $\rho$.

The magnetic moment of $kth-$order, or $(k+1)th$-polar magnetic moment, per atom, can be determined through the rank-$k$ spin nematic tensors $N_{\nu_1 \nu_2 \dots \nu_k}$ as follows~\cite{Kaw.Ued:12}
\begin{equation}
\label{magn.tensor}
M_{\nu_1 \nu_2 \dots \nu_k} \equiv \frac{\Tr \left( \rho^a N_{\nu_1 \nu_2 \dots \nu_k} \right)}{\Tr \rho^a} \, ,
\end{equation}
with $a=n, \, nc$ (or omitted for the total system), and where
\begin{equation}
\label{nematic.tensor}
N_{\nu_1 \nu_2 \dots \nu_k}
\equiv \frac{1}{k!} \sum_{\pi \in P_k}  F_{\pi(\nu_1)} F_{\pi(\nu_2)} \dots F_{\pi(\nu_k)}
\, ,
\end{equation}
here $\pi$ represents the elements of the permutation group of $k$ objects $P_k$. For instance, the dipolar and quadrupolar magnetic moments are simply defined by the operators
\begin{equation}
\label{mag.mom}
N_{\nu} = \, F_{\nu}
\, , 
\quad
N_{\nu_1 \nu_2} = \, \frac{  F_{\nu_1} F_{\nu_2}+ F_{\nu_2} F_{\nu_1}  }{2 } 
\, .
\end{equation}
Recall that the angular momentum operators $F_{\nu}$ are proportional to the tensor operators $T_{1\mu}$~\cite{Var.Mos.Khe:88}
 \begin{equation}
 \label{ang.mom}
    F_z = \cte T_{10} \, , \quad
    F_{\pm} = \mp\sqrt{2} \cte T_{1, \pm 1} \, ,
 \end{equation}
 with $\cte = \sqrt{f(f+1)(2f+1)/3}$ and $F_{\pm} = F_x \pm i F_y$. By addition of the angular momentum operator, and for $k\leq 2f$, the magnetic moment of $kth$-order is a linear combination of the tensor operators $T_{\si \mu}$ with order at most $\si \leq k$ and $|\mu| \leq \Delta$
\begin{equation}
\label{Mag.Tensor}
N_{\nu_1 \nu_2 \dots \nu_k} = \sum_{\si=0}^k \sum_{\mu= -\Delta}^{\Delta} b_{\si \mu} T_{\si \mu} \, ,
\end{equation} 
where $\Delta$ is equal to the total number of $x$ and $y$ subindexes of $N_{\nu_1 \dots \nu_k}$, and $b_{\si \mu}$ are complex numbers. The explicit decomposition can be obtained by using recursively the formula of the product of two tensor operators \cite{Var.Mos.Khe:88}
\begin{align}
T_{l_1m_1} T_{l_2m_2} = & \sum_{l,m} \chi(l_1,l_2,l;f) C_{l_1m_1,l_2m_2}^{lm} T_{lm} \, ,
\end{align}
with 
\begin{align}
\chi(l_1,l_2,l;f) \equiv & (-1)^{2l_2+l-2f}\sqrt{(2l_1+1)(2l_2+1)} 
\nonumber
\\
& \times \sjs{l_1}{l_2}{l}{f}{f}{f}  \,
\label{prod.tens}
\end{align}
written in terms of the 6j-symbol \cite{Var.Mos.Khe:88}.  As an example, we calculate the terms of the 2nd-magnetic moment $N_{\nu_1 \nu_2}$ in Appendix \ref{App.mag}.

The previous results restrict the possible magnetic moments of a condensate $\rho$ with rotational symmetries because, as we will show later, they have several vectors such as $\bm{\rho}_{\si}=0$ for $\si \leq k$. Consequently, some phases of spinor BEC will have isotropic magnetic moments with respect to the physical space, {\it i.e.} $M_{\nu_1 \dots \nu_k} \propto N^{nc} \mathds{1}_{\nu_1 \dots \nu_k}$, where $\mathds{1}_{\nu_1 \dots \nu_k}$ are the components of the identity tensor of $k$ indices. The results are also extended to the different fractions $n$ and $nc$, as well as the total system, because all their respective density matrices are represented in the general expression of $\rho^{nc}$ with a given point group symmetry. 

In the following, we will describe the isotropic $kth$-magnetic moments for the spin-phase families of interest. It is also important to remark that states with vanishing expectation values of the tensor operators $T_{\si \mu}$ are of great relevance in quantum information and quantum metrology. For instance a  property called anticoherence~\cite{Zimba06,Gir.Bra.Bag.Bas.Mar:15}, which in turn is connected to multipartite entanglement~\cite{Bag.Bas.Mar:14,Bag.Dam.Gir.Mar:15}  and the susceptibility is used to detect rotations of quantum systems~\cite{Chr.Her:17,Bou.etal:17,martin2020optimal}. A useful result that we are going to use later is that the anticoherence order of a spin state is equal to its maximal order of homogeneous $kth$-magnetic moment~\cite{Gir.Bra.Bag.Bas.Mar:15}, since, by definition, a spin state $\rho$ is anticoherent of order-$k$ if its $\rho_{\si \mu}$-components are zero for any $0<\si\leq t$ and $|\mu| \leq \si $. 
%
%
%
%
%
%
%
%
\subsection{ $\ket{f,m}$ phases}
The simplest family we address in this work is the  set of $F_z$-eigenvectors $\ket{f,m}$, with spinor order-parameter $\phi_k \propto \delta_{km}$. The family contains all the states with a continuous point group $SO(2)$, equivalent to rotations about the z axes. The case $m=0$ has the time-reversal operator as an additional symmetry. Thus the corresponding $\rho^{nc}$ of this family would be diagonal over the $\{\ket{f,m} \}$ basis
\begin{equation}
\label{mix.fm}
\rho^{nc} = \sum_{m=-f}^f \la_m \ket{f,m}\bra{f,m} \, .
\end{equation}
In terms of the Majorana representation of mixed states, the components of the $\bm{\rho}_{\si}$ vectors are
\begin{equation}
\label{Mag.fm}
\rho_{\si \mu} = r_{\si} \delta_{\mu 0} \, ,
\end{equation}
 where the norms of the vectors are linear combinations of the eigenvalues $\la_m$. The constellations of $\con{\rho^{nc}}$ have $\si$ points on each pole (see Fig.~\ref{MSC.FFNOON}). The additional symmetry associated to the $\ket{f,0}$ spin-phase yields that $\la_m = \la_{-m}$, consequently the $\bm{\rho}_{\si}$ for $\si$ odd are zero. Some examples of the $\ket{f,m}$-phases are the FM and P phases of a spin-2 BEC, which are proportional to the $\ket{2,2}$ and $\ket{2,0}$ spin states, respectively. 

The eigenspectrum \eqref{mix.fm} of $\rho^{nc}$ implies that the only non-vanishing terms in $M_{\nu_1 \dots \nu_k}$ are given by $T_{\si 0}$ terms. For instance, the first two magnetic moments are given by
\begin{align}
M_{\nu} = \delta_{\nu z} \sum_{m=-f}^f m \La_m = \delta_{\nu z} \cte N^{nc} \rho_{1 0}
\, ,   
\end{align}
and
\begin{align}
    M_{xy}=&M_{xz} = M_{yz} =0 \, ,
\nonumber
    \\
    M_{zz} = & \frac{f(f+1) N^{nc}}{3} +\frac{1}{6\sqrt{5}}\sqrt{\frac{(2f+3)!}{(2f-2)!}} \rho_{20} \, ,
    \\
    M_{xx} = &M_{yy} = \frac{f(f+1) N^{nc}}{3} -\frac{1}{12\sqrt{5}}\sqrt{\frac{(2f+3)!}{(2f-2)!}} \rho_{20} \, , 
    \nonumber
\end{align}
where we use the equations deduced in App.~\ref{App.mag} and the fact that $T_{00} = \mathds{1} / \sqrt{2f+1}$. In particular, let us notice that the $\ket{f,0}$ spin phase, equivalent to the polar phase, must have $M_{\nu}=0$.
\subsection{NOON spin phases}
The family consists of the spin-phases $\bm{\Phi}$ with components of the form
\begin{equation}
\label{Noon.po}
\phi_{m} \propto ( \delta_{f,m} + \delta_{-f,m}) / \sqrt{2} \, ,
\end{equation}
that we define as the NOON spin-phases since they are equivalent of a quantum superposition of $2f$ spin-1/2 states pointing about orthogonal directions~\cite{agarwal12}. The square (S) spin-2 phase is an example of the NOON spin phase. Its point group is equal to $D_{2f}$ in the Schönflies notation \cite{book.bra.cra:10}, which is equivalent to the point group associated to the regular $2f$-agon. The coefficients of $\bm{\rho}_{\si}$ of the noncondensate fraction are 
\begin{equation}
\label{Mag.Sq}
\rho_{\si \mu} = \left\{ 
\begin{array}{c c}
r_{\si} \delta_{\mu 0} & \text{ if } \si < 2f, \si \text{ even}
\\
r_{2f} \left( \sin x \, \delta_{\mu 0} + \frac{\cos x}{\sqrt{2}}\delta_{\mu,\pm 2f } \right)
& 
\text{ if } \si = 2f
\\
0 & \text{ otherwise}
\end{array}
\right. \, .
\end{equation}
Thus, the constellations are conformed by all the $\si$ points in each pole for $\si < 2f$, and by a regular $n$-agon prism for $\si=2f$, where its height is dependent of the variable $x$. Consequently, the $\bm{\rho}_{\si}$ vectors of the NOON and $\ket{f,0}$ spin phases are equal except for $\si=2f$ (see Eqs.~\eqref{Mag.fm}-\eqref{Mag.Sq}). This implies that the $k$th-magnetic moments $M_{\nu_1 \dots \nu_k}$ (Eq.~\eqref{Mag.Tensor}) of $\rho^{nc}$ of the NOON  and $\ket{f,0}$ spin phases have the same expressions for $k < 2f$. Therefore, assuming that just the magnetization of the condensate is being measured, the NOON and the $\ket{f,0}$ spin phases can only be distinguished by its $(2f)$th-magnetic moment.

Favorably,  the eigenspectrum $(\la_{\nu} , \bm{v}_{\nu})$ of the NOON phase can be diagonalized analytically for a general spin value $f$. It includes $(f-1)$ pairs of degenerate eigenvectors, $\bm{v}_{2k-1}$ and $\bm{v}_{2k}$ with $\la_{2k-1}=\la_{2k}$ and where $k$ goes from 1 to $(f-1)$. The components of the eigenvectors $v_{\nu,m}$ are given by
\begin{align}
 v_{1 ,m} = &\, \delta_{m ,f-1 } \, , & v_{2, m } = & \, \delta_{m, -f+1 }  \, ,
\nonumber
\\
v_{3, m} = & \, \delta_{m, f-2 } \, , & v_{4, m } = & \, \delta_{m, -f+2 } \, ,  
\nonumber
\\
\vdots & &  \vdots &
\nonumber
\\
 v_{2f-3, m} = & \, \delta_{m, 1 } \, , & v_{2f-2, m } = & \, \delta_{m, -1 } \, .
\label{eig.Sq1}
\end{align}
There are also 3 eigenvectors not necessarily degenerate,
\begin{align}
v_{2f-1, m} = &\, \delta_{m0} \, , 
\nonumber
\\ 
v_{2f, m} = & \, \frac{1}{\sqrt{2}}\left( \delta_{m,-f} - \delta_{m,f}  \right) \, , 
\nonumber
\\ 
v_{2f+1, m} = & \, \frac{1}{\sqrt{2}}\left( \delta_{m,-f} + \delta_{m,f}  \right) \, , 
\label{eig.Sq2}
\end{align}
where the last eigenvector is proportional to the order parameter $\bm{\Phi}$ \eqref{Noon.po}. 
%
%
%
\def\sc{1.95cm}
\def\hc{1cm}
\begin{figure}[t!]
\begin{tabular}{|c|c|}
\hline
 $\mathcal{C}_{\bm{\Phi}} $ &
$\mathcal{C}_{\rho^{nc}}$
\\
\hline
\begin{tabular}{c}
 T ($f=2$)
\\
\includegraphics[width=\sc]{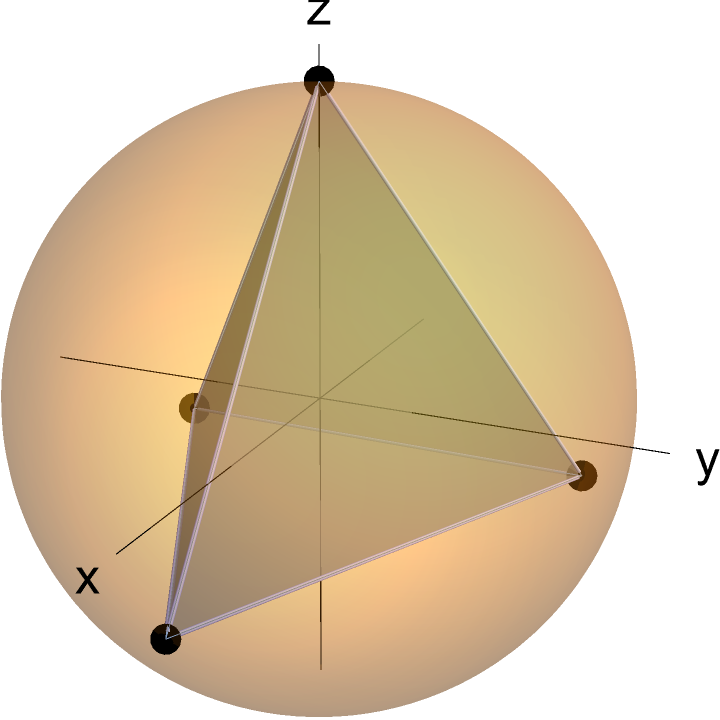}
\end{tabular}
&
\begin{tabular}{c}
$\si=3$ \hspace{\hc} $\si=4$
\\
\includegraphics[width=\sc]{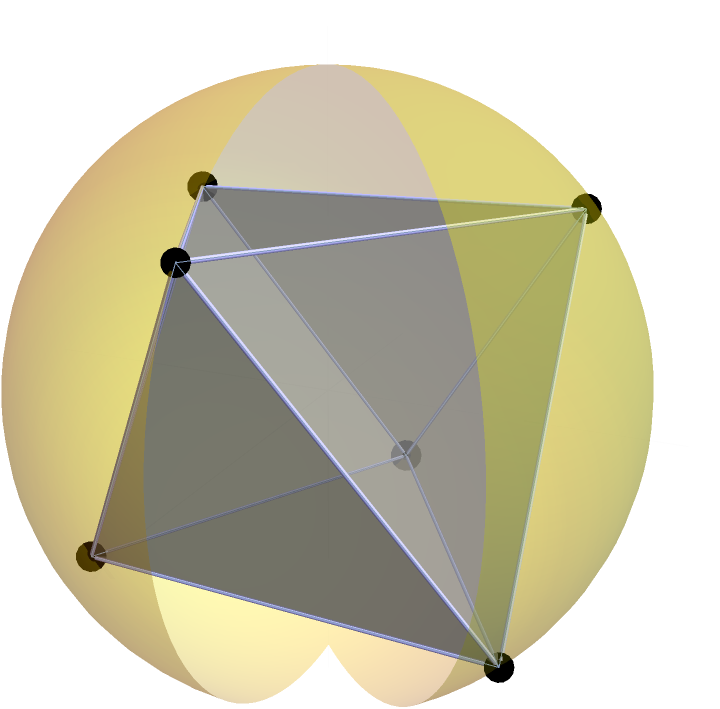}
\includegraphics[width=\sc]{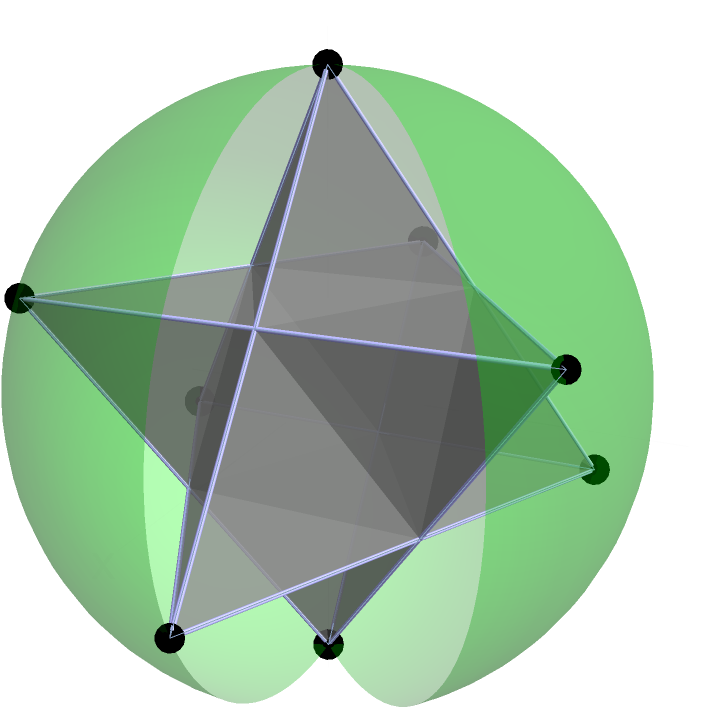}
\end{tabular}
\\
\hline
\begin{tabular}{c}
O ($f=3$) 
\\
\includegraphics[width=\sc]{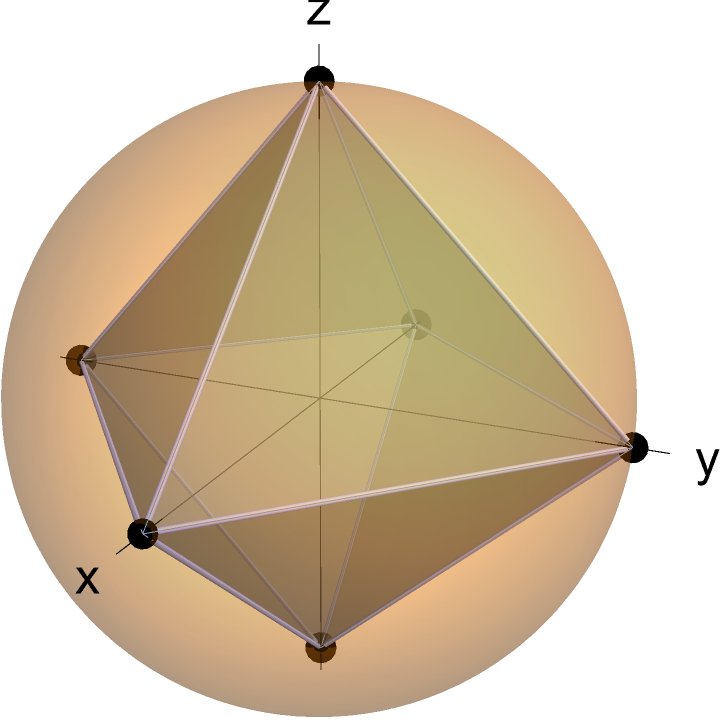}
\end{tabular}
& 
\begin{tabular}{c}
$\si=4$ \hspace{\hc} $\si=6$
\\
\includegraphics[width=\sc]{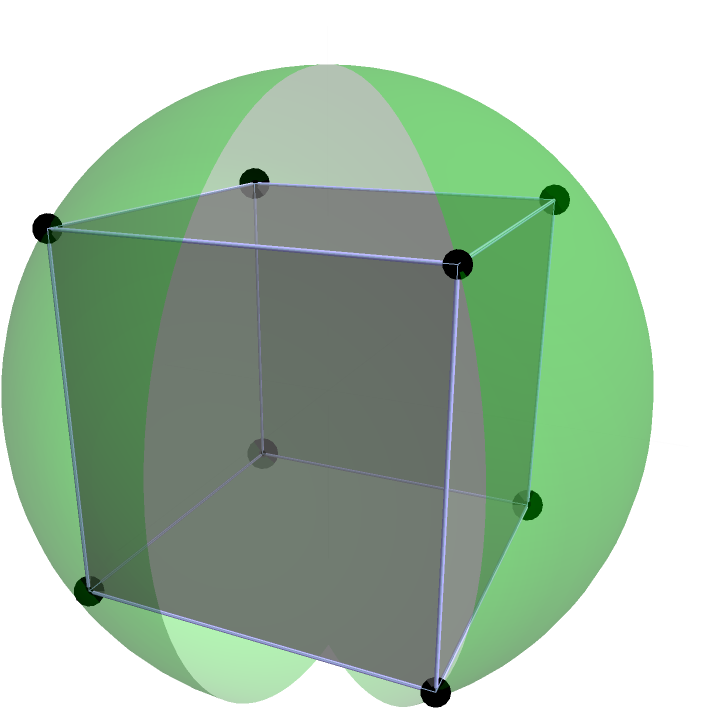} 
\includegraphics[width=\sc]{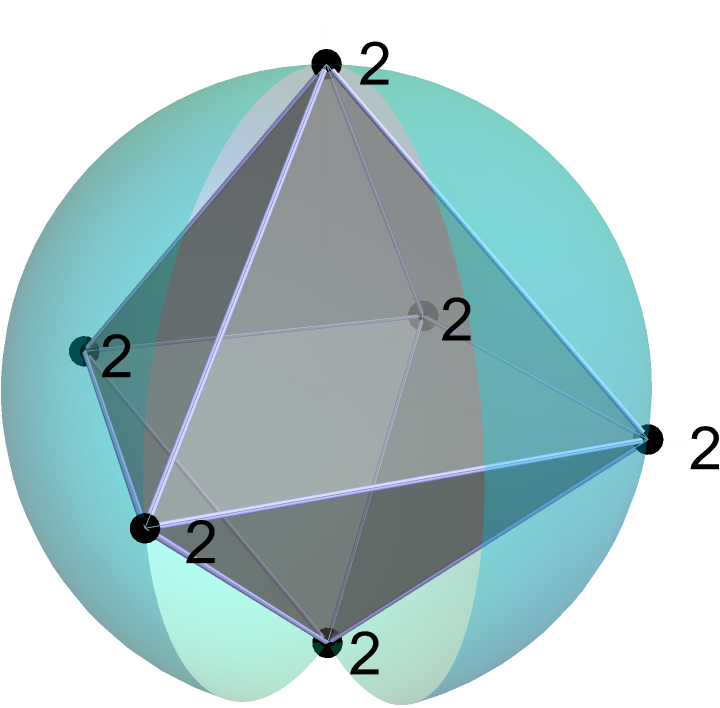}
\end{tabular}
\\
\hline
\begin{tabular}{c}
C ($f=4$) 
\\
\includegraphics[width=\sc]{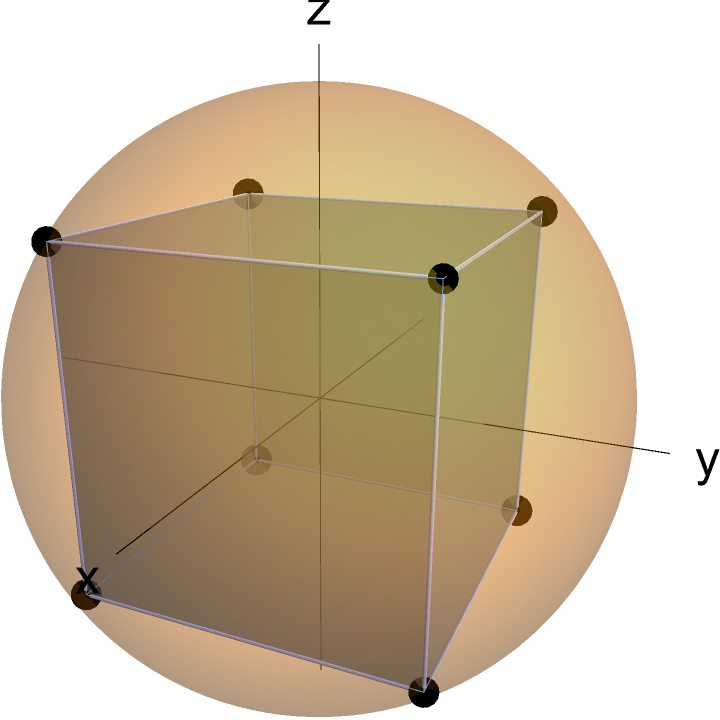} 
\end{tabular}
&
\begin{tabular}{c}
$\si=4$ \hspace{\hc} $\si=6$ \hspace{\hc} $\si=8$
\\
\includegraphics[width=\sc]{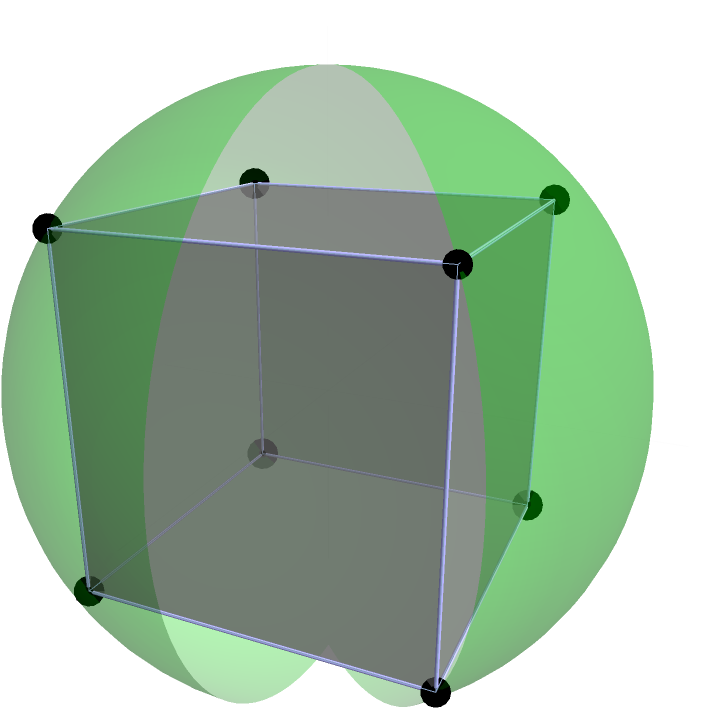}
\includegraphics[width=\sc]{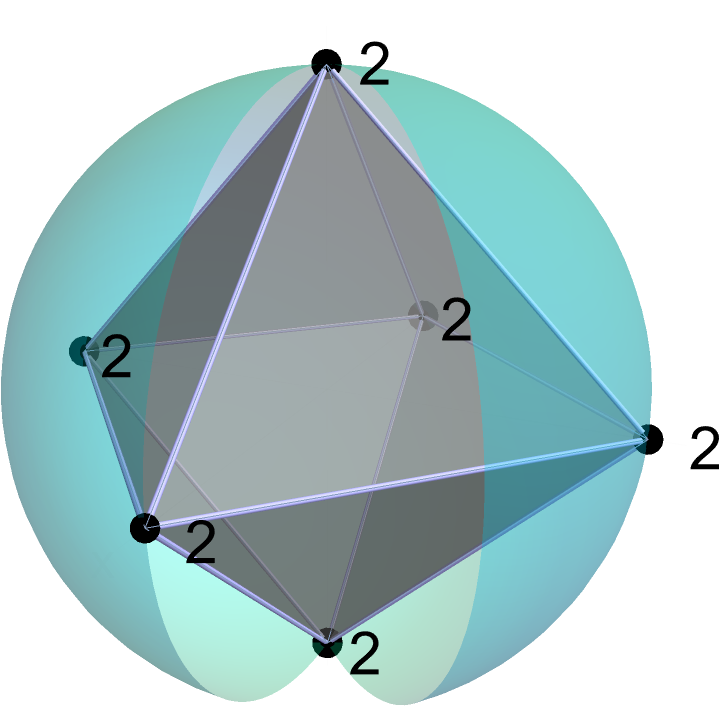}
\includegraphics[width=\sc]{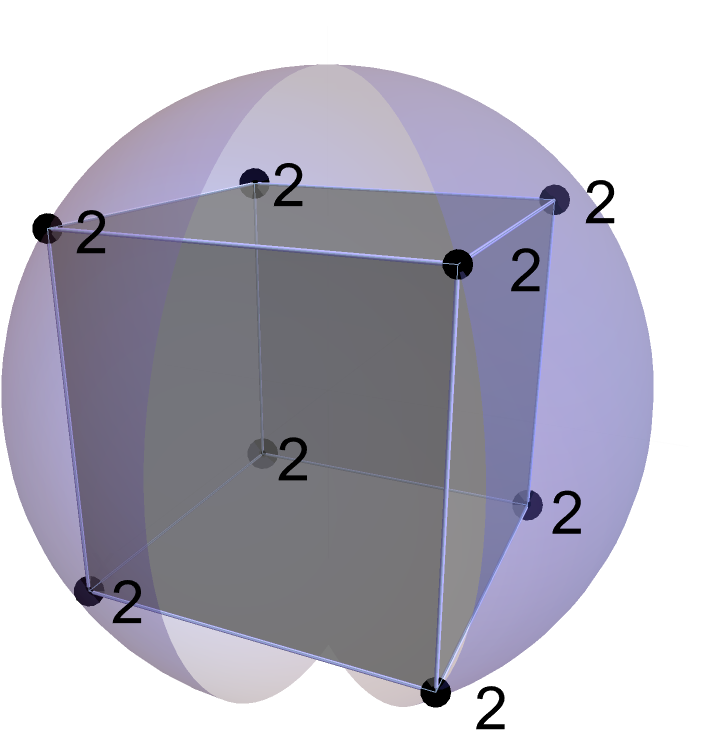}
\end{tabular}
\\
\hline
\begin{tabular}{c}
I ($f=6$)
\\
\includegraphics[width=\sc]{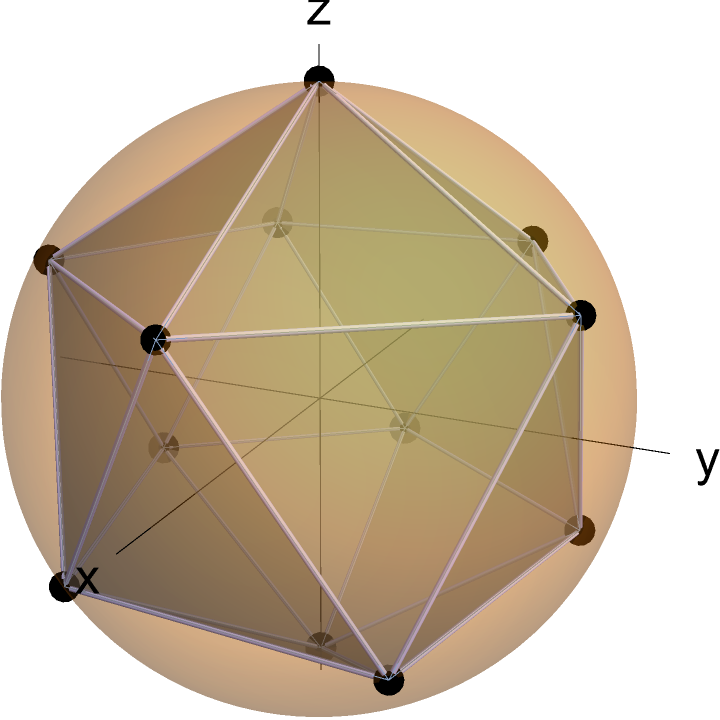} 
\end{tabular}
& 
\begin{tabular}{c}
$\si=6$ \hspace{\hc} $\si=10$ \hspace{\hc} $\si=12$
\\
\includegraphics[width=\sc]{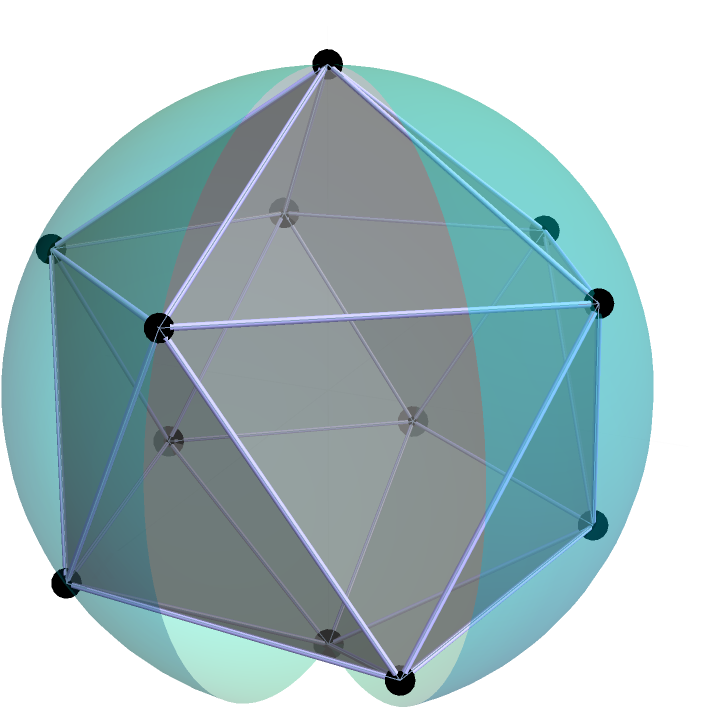}
\includegraphics[width=\sc]{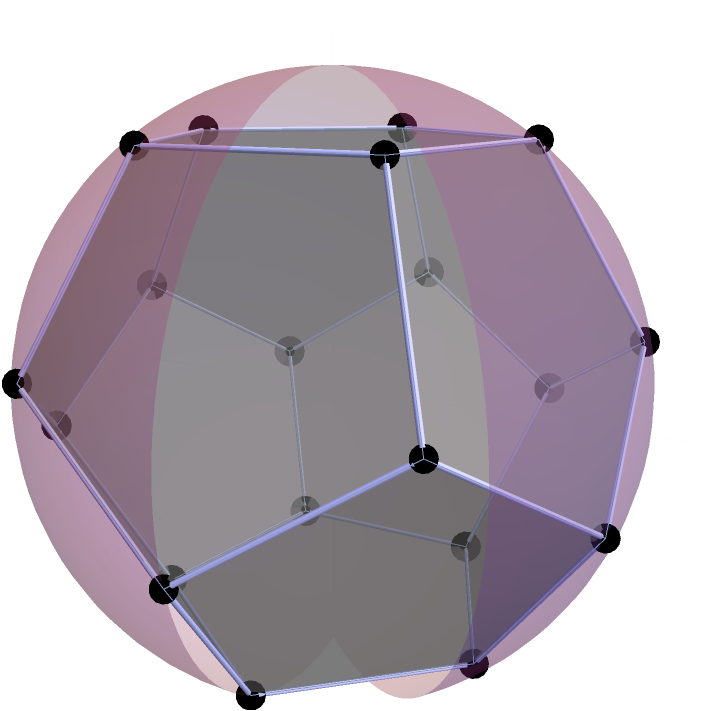}
\includegraphics[width=\sc]{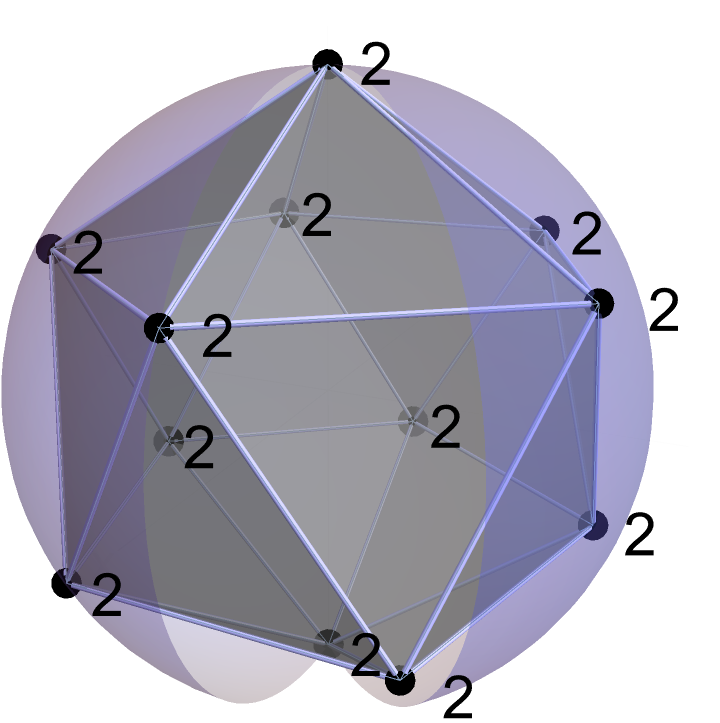}
\end{tabular}
\\
\hline
\end{tabular}
\caption{\label{MSC.Platonic} Majorana representation of mixed states with a point group equal to a platonic solid: tetrahedron (T), octahedron (O), cube (C) and icosahedron (I). The radii of the spheres of $\con{\rho^{nc}}$ are arbitrary up to the semi-definite condition of $\rho^{nc}$.
} 
\end{figure}
\subsection{Platonic phases}
Spin-phases of BEC with point group symmetry  corresponding to that of the platonic solids $-$tetrahedron (T), octahedron (O), cube (C), and icosahedron (I)$-$ appears for condensates with spin $f=2 ,3,4$ and $6$, respectively~\cite{Kawa.Phuc.Blakie:2012,PhysRevA.84.053616}. Here, we shall not characterize the spin phase of $\rho^{nc}$ having the point group symmetry of the dodecahedron, corresponding to a BEC spin phase with $f=10$. In any case, it is noteworthy that a spin-10 BEC has yet to be developed in laboratory. We now proceed to enlist the properties of each platonic phase of interest, along with its corresponding non-zero components of the $\rho_{\si \mu}$ vectors:
%

(1) Tetrahedron (T) phase:  
\begin{align}
 \rho_{\, 3 \mp 3}  = & \pm \frac{\sqrt{2}}{3} \, , \quad
&
\rho_{4 \mp 3} = & \pm \sqrt{\frac{10}{27}} \, ,
\nonumber
\\
\rho_{\, 30} = & \frac{\sqrt{5}}{3} \, , \quad
& 
\rho_{40} = & -\sqrt{\frac{7}{27}} \, .
\label{Tet.cont}
\end{align}
The constellations $\con{\rho^{nc}}$ form an octahedron and a pair of antipodal tetrahedrons, respectively. It is easy to demonstrate that such constellations have the symmetry $T$, since the point group $T$ belongs to the point group of the octahedron $O$. Due to the vanishing $\bm{\rho}_{\si}$-vectors of $\si=1,2$ for the T phase, the first two magnetic moments are isotropic, implying that
\begin{equation}
M_{\nu} = 0 \, , \quad M_{\nu_1 \nu_2} = 2 \delta_{\nu_1 \nu_2}  N^{nc} \, .
\end{equation}
For a lower spin $f=1$, a phase with this magnetization is forbidden. Hence, it can be said that the T-phase is the first spin phase with its first two multipolar magnetic moments isotropic. To obtain isotropic magnetic moments of higher order, one must consider a spin BEC of higher spin. 

Let us now show that the noncondensate fraction $\rho^{nc}$ of the T-phase has only three degrees of freedom $(N^{nc},r_3,r_4)$. This reduction allows us to easy calculate the eigenspectrum of $\rho^{nc}$ analytically, since its characteristic polynomial, of degree 5, leads to two non-degenerate roots, and a set of three-fold degenerate roots. The exact eigenspectrum is given by
\begin{align}
\La_1 = & \La_2 = \La_5 = \frac{1}{5} - \sqrt{\frac{2}{15}} r_4  \, ,
\nonumber
\\
\La_{3,4} = & \frac{1}{10} \left( 2 \mp 5 \sqrt{2} r_3 + \sqrt{30} r_4 \right) \, .
\nonumber
\\
\bm{v}_{1} = & (0,0,1,0,0)^{\text{T}} \, , 
\nonumber
\\
\bm{v}_{2}= & (0,1,0,0,\sqrt{2})^{\text{T}}/\sqrt{3}
\, , \quad
\bm{v}_{3}= (0,-\sqrt{2},0,0,1)^{\text{T}}/\sqrt{3} \, ,
\nonumber
\\
\bm{v}_{4}= & (1,0,0,\sqrt{2},0)^{\text{T}}/\sqrt{3}
\, , \quad
\bm{v}_{5}= (-\sqrt{2},0,0,1,0)^{\text{T}}/\sqrt{3} \, .
\end{align}

The density matrix $\rho^{nc}$ is a physical state (with non-negative eigenvalues) if the variables $(r_3,r_4)$ fulfill the condition $5|r_3 |\leq \sqrt{2} + \sqrt{15} r_4$, with $r_3 \in \left[ -\frac{1}{\sqrt{2}},\frac{1}{\sqrt{2}} \right]$ and $r_4 \in \left[ - \sqrt{\frac{2}{15}} \sqrt{\frac{3}{10}} \right]$. 
%
%
%
%

(2) Octahedron (O) phase:
\begin{align}
\rho_{4 \pm 4} =& \sqrt{\frac{5}{24}} \, , \quad
&
\rho_{6 \pm 4} =& -\sqrt{\frac{7}{16}} \, ,
\nonumber
\\
\rho_{40} =& \sqrt{\frac{7}{12}} \, , \quad
&
\rho_{60} =& \frac{1}{\sqrt{8}} \, ,
\label{Oct.cont}
\end{align}
with constellations describing a cube for $\si=4$, and an octahedron for $\si=6$ with all its stars having degeneracy equal to 2. The cube and the octahedron belong to the same point group symmetry, the alternate group $A_4$~\cite{book.bra.cra:10}, as  they are dual geometrical figures. The $O$-phase is the first spin-phase with its first three magnetic moments isotropic 
\begin{equation}
M_{\nu}= M_{\nu_1 \nu_2 \nu_3}=0 \, ,
\quad
M_{\nu_1 \nu_2} = 4N^{nc}\delta_{\nu_1 \nu_2} \, ,
\end{equation}
since $\bm{\rho}_{\si} = 0$ for $\si=1,2,3$. The proof that the nonexistence of a spin phase with the same property for $f\leq 2$ comes from the fact that there are no spin-$f$ states with anticoherence of order 2 for $f<2$~\cite{Bag.Dam.Gir.Mar:15}. 

%
(3) Cube (C) phase:
\begin{align}
& & & & \rho_{8 \pm 8} = & \sqrt{\frac{65}{384}} \, ,
\nonumber
\\
\rho_{4 \pm 4} = & \sqrt{\frac{5}{24}} \, , 
\quad
&
\rho_{6 \pm 4} = & -\frac{\sqrt{7}}{16} \, , \quad
&
\rho_{8 \pm 4} = & \sqrt{\frac{7}{96}} \, ,
\nonumber
\\
\rho_{40} = & \sqrt{\frac{7}{12}}  \, ,
&
\rho_{60} = & \frac{1}{\sqrt{8}} \, ,
&
\rho_{80} = & \frac{\sqrt{33}}{8} \, .
\label{Cub.cont}
\end{align}
Then $\con{\rho^{nc}}$ has three constellations conformed by a cube for $\si=4$ and $8$, and by an octahedron for $\si=6$. The stars of the constellations of $\si=6$ and 8 are two-fold degenerate. Similarly as the O phase, the C phase has isotropic magnetic moments of order $\si=1,2,3$.

(4) Icosahedron (I) phase:
\begin{align}
& 
& \rho_{10 \, \pm 10} = \sqrt{\frac{187}{1875}}  \, , \text{ } 
& \rho_{12 \, \pm 10} = \sqrt{\frac{741}{3125}}
\, , 
\nonumber
\\
& \rho_{6 \pm 5}  = \pm  \frac{\sqrt{7}}{5} \, ,
& \rho_{10 \pm 5} = \pm \frac{\sqrt{209}}{25} \, , \text{ } 
& \rho_{12 \pm 5} = \mp \sqrt{\frac{286}{3125}} \, ,
\nonumber
\\
& \rho_{60} = - \frac{\sqrt{11}}{5} \, ,
& \rho_{10 \, 0} = \sqrt{\frac{247}{1875}} \, ,
\text{ } 
& \rho_{12 \, 0} = 3\sqrt{\frac{119}{3125}} \, ,
\label{Ico.cont}
\end{align}
Here, as it is expected, the constellations are conformed by icosahedrons and dodecahedrons because they are dual polyhedra with point group isomorphic to the alternate group $A_5$. The stars of the constellation of $\si=12$ are doubly-degenerated. 
The Majorana representation of the I phase tells us that the first five multipolar magnetic moments are isotropic. The I phase is the first spin value with this property since there are no anticoherent spin states of order 5 for spin values $f<6$~\cite{Bag.Mar:17}.
\section{Finite temperatures}
\label{Sec.Finite}
\remaRe{Once we have characterized the possible noncondensate fractions $\rho^{nc}$ with a particular point group symmetry, we can use them to study a spinor BEC condensate in a model with self-consistent symmetries.} In this section we derive within the HF approximation the equations that define the temperature dependent condensate and noncondensate fractions, $\rho^c$ and $\rho^{nc}$, of a spin-2 BEC. Next, we exploit the results discussed in the previous section that will allow us to reduce significantly  the calculations and yields to analytical expressions of the some physical properties of the condensate.
\subsection{HF equations}
We start by writing explicitly the HF energy of a spin-2 BEC by using the equations Eqs.~\eqref{c0.HF}-\eqref{c2.HF} containing all the spin-interactions
\begin{align}
& E_{HF} = E_s + \frac{c_0}{2} 
\left\{ 
N^2 + \Tr \left[ \rho^{nc} \left( 2\rho^c + \rho^{nc} \right) \right]
\right\}
\nonumber
\\
& + \frac{c_1}{2} \sum_{\alpha} \left\{
\Tr \left[ \rho F_{\alpha} \right]^2 + \Tr[F_{\alpha} \rho^{nc} F_{\alpha} (2\rho^c+ \rho^{nc})]
\right\}
\nonumber
\\
& + \frac{c_2}{10} \left\{ \Tr \left[
\TRS \rho \TRS \rho + \TRS \rho^{nc} \TRS \left( 2\rho^c + \rho^{nc} \right)
\right] \right\} - \mu \left( \Tr \rho - N \right) \, .
\end{align}
The new two-body interactions with respect to the MF energy are the direct and exchange interactions, respectively. Here, \remaRe{$E_{s}$ is the spatial energy associated to $h_s$ in Eq.~\eqref{Full.Ham}. In our case $U(\bm{r})=0$, the spatial eigenstates are labeled by the wavevector $\bm{k}$, and their corresponding eigenenergies are $E_s = \hbar^2 k^2/2M$, {\it i.e.} just the kinetic energy}. The effect of each term over the spin coherence and the distribution of the atoms in the magnetic sublevels is discussed in \cite{Kawa.Phuc.Blakie:2012}.

The condensate fraction of the system $\rho^c=N^c \bm{\Phi} \bm{\Phi}^{\da}$ is a pure state with $\bm{k}=\bm{0}$. Hence, the resulting GP equations that results from minimizing $E_{HF}$, ($\delta E_{HF}/\delta \phi^*_m =0 $), are given by a system of three (non-linear) equations involving $\phi_m$ and $\rho^{nc}$. On the other hand, $\rho^{nc}$ is written as a sum of its eigenvectors $\bm{\xi}^{\la} = (\xi_f^{\la}, \xi_{f-1}^{\la}, \dots, \xi_{-f}^{\la})^T$ weighted by its Bose-Einstein distribution factor $n_{\la}$, 
\begin{equation}
\rho^{nc}_{ij} = \sum_{\la} n_{\la} \xi^{\la}_i \xi^{\la*}_j  
\, , \quad
n_{\lambda} =\left( e^{\beta \epsilon_{\la}} -1 \right)^{-1}
\, .
\end{equation}
 The global subindex $\la$ includes the spatial and spinor quantum numbers, $\la= (\bm{k}, \nu)$, with $\nu=1,2,\dots, 2f+1$ and $\beta= 1/k_{B}T$ where $k_{B}$ is the Bolztmann constant. The eigenvectors $\bm{\xi}^{\la} $ and their associated energies $\epsilon_{\la}$ are obtained by the noncondensate Hamiltonian $A$, given by $A_{ij} = \delta E_{HF}/\delta \rho^{nc}_{ji} $. The decoupling of the spatial and spinor parts in the Hamiltonian $A$ leads to $\epsilon_{\lambda}= -\hbar^2 k^2/2M + \kappa_{\nu}$, with $\kappa_{\nu}$ the eigenvalue of the spinor part of $A$. \remaRe{In summary, $\rho^{nc}$ is an statistical mixture of thermal atoms with eigenstates labeled by the wavevector $\bm{k}$, and the index $\nu$ that specifies the spinor eigenstate of the HF Hamiltonian $A$.} The spatial part of $\rho^{nc}$ can be integrated using that $\sum_{\bm{k}}\rightarrow (2\pi)^{-3} \int\diff \bm{k}$ \footnote{\remaRe{The full spatial density matrix of $\rho^{nc}$ includes its off-diagonal terms $\langle \delta^{\dagger}(\bm{r}_1) \delta(\bm{r}_2)\rangle$. However, since we are only interested in the spinorial sector, we integrate over the space, and only the diagonal terms contribute in Eq.~\eqref{Poly.sta}.}},
\begin{equation}
\label{Poly.sta}
\rho^{nc}_{ij} = \sum_{\nu=1}^{2f+1} \xi^{\nu}_i \xi^{\nu *}_j \La_{\nu} \, , \quad \La_{\nu} =  \frac{Li_{3/2}\left(e^{-\beta \kappa_{\nu}} \right)}{\la_{dB}^3}  \, ,
\end{equation}
where $Li_{3/2}(z)$ is the polylogarithm function and $\la_{dB} = h / \sqrt{2\pi M k_B T} $ is the thermal de Broglie wavelength. The eigendecomposition of $A_{ij}$, which is now a $(2f+1)\times (2f+1)$ matrix, are called the HF equations.  The atom fractions $N^c$ and $N^{nc}$ can be written in terms of $\ka_{\nu}$ because $N^{c}=N-N^{nc}$ and $N^{nc}= \sum_{\nu} \La_{\nu}$. Moreover, the $A$ matrix, dependent of $\rho^c$ and $\rho^{nc}$, can also be written in terms of the $\Lambda_{\nu}$ and then of the $\ka_{\nu}$ variables. Finally, we use the fact that $A$ and $\rho^{nc}$ share the same eigenvectors, leading us to obtain a system of algebraic-transcendental equations for the $\ka_{\mu}$ eigenenergies of $A$. 

The equations to determine $\rho^c$ are still called Gross-Pitaevskii (GP) equations $\delta E_{HF}/\delta \phi^*_m =0 $, and they can be written as $\mu \bm{\Phi} = L \bm{\Phi}$,  with
\begin{align}
L = & \, c_0 \big( N \mathds{1}_f + \rho^{nc} \big)
 + c_1  \sum_{\al} \Big\{ \Tr \left[F_{\al} \rho \right] F_{\al} + F_{\al} \rho^{nc} F_{\al} \Big\} 
\nonumber
\\ &
+ \frac{c_2}{5} \TRS \left( \rho + \rho^{nc} \right) \TRS  \, ,
\label{HF.GPE}
\end{align}
where $\al = x,y$ and $z$. On the other hand, the noncondensate Hamiltonian $A$, with $A_{ij} = \delta E_{HF} / \delta \rho_{ji}^{nc} $ that allows to determine $\rho_{nc}$, are called the Hartree-Fock (HF) equations, and is explicitly given by
\begin{align}
A = & L -\mu \mathds{1}_5 +c_0 \rho^c + c_1 \sum_{\al} F_{\al} \rho^{c} F_{\al} + \frac{c_2}{5} \TRS \rho^c \TRS  
\nonumber
\\
= & -\mu \mathds{1}_5 + c_0 \big( N \mathds{1}_5 + \rho \big)
\nonumber
\\ &
 + c_1 \sum_{\al} \Big\{
 \Tr \left[ \rho F_{\al} \right] F_{\al} + F_{\al} \rho F_{\al} \Big\}
+ \frac{2c_2}{5} \TRS \rho \TRS  \, .
\label{HF.nc}
\end{align}
The standard procedure is to solve in a self-consistent fashion the GP-HF equations \eqref{HF.GPE}-\eqref{HF.nc}~\cite{blaizot1986quantum,griffin2009bose,Kawa.Phuc.Blakie:2012}. However, as it is shown below, the symmetries inherited by $\rho^{nc}$ reduce considerably its degrees of freedom and therefore the complexity of the problem. Moreover, the case spin-2 BEC is amenable for the calculation of the associated eigenvectors of each $\rho^{nc}$, as well as for the derivation of closed equations for the $\ka_{\nu}$ energies. In the next section we discuss the calculation of the expressions of $\ka_{\nu}$ for low temperatures, and analize the admissible regions of each allowed spin phase. Also, the phase diagrams at finite temperatures can be calculated by finding the ground states defined in terms of the thermodynamic HF potential, $\Phi_{HF} = E_{HF} - T S_{HF}$, where $S_{HF}$, is the HF entropy \cite{blaizot1986quantum,Ser.Mir:21}. 

\begin{figure}[t!]
\includegraphics[scale=0.467]{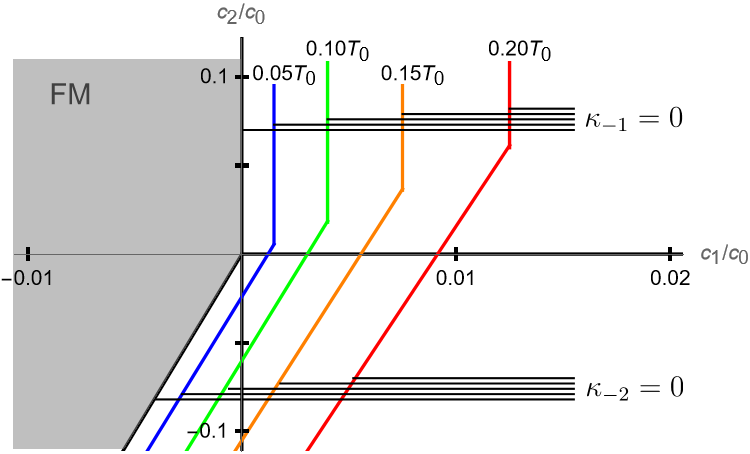}
\\
\includegraphics[scale=0.467]{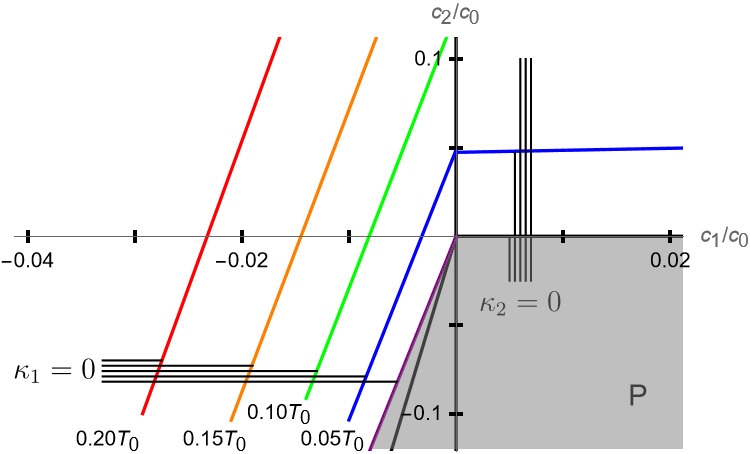}
\\
\includegraphics[scale=0.467]{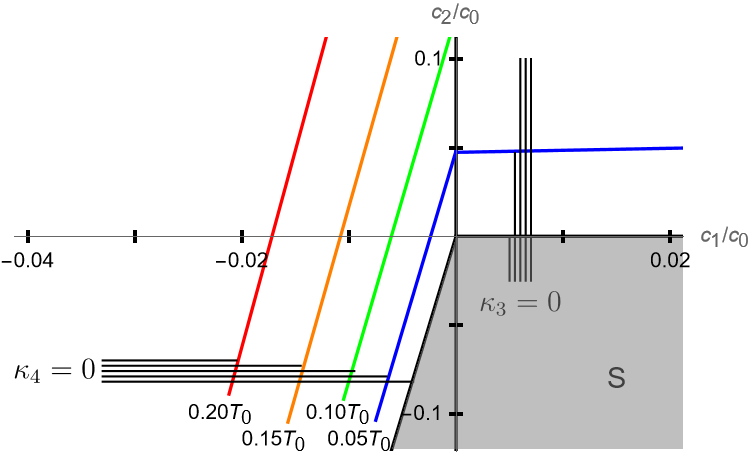}
\\
\includegraphics[scale=0.467]{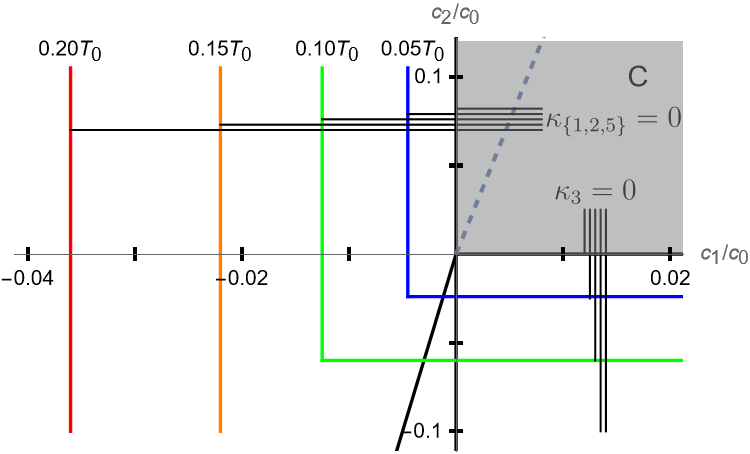}
\caption{\label{Adm.region} 
Allowed regions of the spin-2 phases of a BEC in the $(c_1 ,c_2)$ parameter space at temperatures $T/T_0=0,0.05,0.1,0.15,0.2$, respectively. The black solid lines are the ground-phase boundaries. The regions where each phase exists at $T=0$ is remarked with the gray region, which coincides also with the zone where it is the ground state except for the S phase. The physical nature of each phase boundary is also mentioned in the figure. The symbol $\kappa_{\{1 \, , 2 \, , 5 \}}=0$ in the C phase means that the three eigenenergies $\kappa_{1} \, , \kappa_{2}$ and $\kappa_{5}$ vanish. The dashed line corresponds to the $(c_1,c_2)-$values where $\rho^{nc}$ has a fourth-degenerated eigenvalue. } 
\end{figure}
\subsection{Admissible regions of the phases}
Following the HF approximation, the $\kappa_{\nu}$ energy is by definition the additional energy that an atom required to be added in $\rho^{nc}$ in the spin state $\bm{\xi}_{\nu}$ with order-parameter $\bm{\Phi}$~\cite{Ser.Mir:21}. Hence, the energies must satisfy that
\begin{equation}
\label{HF.kappa}
\kappa_{\nu}>0 \, ,
\end{equation}
otherwise the $\bm{\Phi}$ phase becomes forbidden and may promotes a drastic (quench) evolution to another phase. The condition \eqref{HF.kappa} gives us valuable insights about the regions in which spin phase could exist. In particular, an scenario in which regions of different spinor phases overlaps (coexist), can also occur~\cite{Ser.Mir:21}. While the spin phase with the lowest thermodynamic potential gives us the ground phase of the BEC, the others phases give rise to metastable phases~\cite{Ser.Mir:21}. The characterizations of the noncondensate fractions $\rho^{nc}$ presented in the previous sections allow us to calculate the analytical expressions of all the $\kappa_{\nu}$ energies. In what follows, we discuss how we proceed to calculate the allowed regions in the $(c_1 , c_2)$ space, as well as on the physical nature of their boundaries for the case of a spin-2 BEC. For simplicity, we write the equations in a compact form, while the full expressions and details are provided in Appendix \ref{spin2.kappa}. 
The new admissible region of each phase, appearing as we increase the temperature (see Fig.~\ref{Adm.region}), includes a region where the phase is metastable~\cite{Ser.Mir:21}. \remaRe{We also obtained that, while for $T=0$, the whole nematic family is equally valid to be a ground phase, for $T\neq0$ only the square and polar phases remain as ground states, which are in fact the states with higher point group in the nematic family. This is called in the literature order-by-disorder phenomenon and it has been predicted for the nematic family of spin-2 BEC in Refs.~\cite{PhysRevLett.98.190404,PhysRevLett.98.160408}. }
\subsubsection{FM case}
%
%
Let us consider the FM phase of a spin-$f$ BEC, oriented along the $z$ axes~\footnote{This could be achieved by a process of constant magnetization about the $z$ axes, and then turning the fields off $p=q=0$.}, and described in general by a state  $\ket{f,f}$. The degrees of freedom of a generic noncondensate fraction $\rho^{nc}$ of spin $f$ is $(2f+1)^2$, being 25 in the particular case of $f=2$. However, exploting its symmetries, they are reduced to just 5 for the FM case, which are the $\La_m$ variables of $\rho^{nc}= \sum_{m=-2}^2 \La_m \ket{2,m} \bra{2,m}$. Moreover, since the atom fractions $N^c$ and $N^{nc}$ can be written in terms of $\La_m$ because $N^{c}=N-N^{nc}$ and $N^{nc}= \sum_{m} \La_m$,  the whole problem has only five unknown variables $\La_m$. These variables are functions of the eigenenergies of $A= \delta E_{HF} / \delta \rho^{nc}_{ji}$, $\ka_{m}$ \eqref{Poly.sta}. In summary, the  eigenvalues of $A$ leads us to a set of equations for the $\ka_{\nu}$ energies in terms of a linear combination of the $\La_{\nu}$, which are written explicitly in App.~\ref{spin2.kappa}. At low temperatures and with $N\sim 10^{4}$ atoms per $cm^{3}$, which is a typical value in the experiments, the terms in the energies $\ka_{\nu}$ proportional to $N$, becomes the dominant terms, defining the physical boundaries of the allowed regions. For the FM phase, such boundaries are constrained by the conditions $\ka_{-1}=0$ and $\kappa_{-2} = 0$ (see Fig.~\ref{Adm.region}), where 
\begin{align}
\ka_{-1}^{(FM)} = & -6c_1 N + \mathcal{O}(\La_{\nu})
\, , 
\\
\ka_{-2}^{(FM)} = & \, 2N \left(-4 c_1 + \frac{c_2}{5} \right) + \mathcal{O}(\La_{\nu})
\, .
\end{align}
We can observe that when $T=0$, and then $\La_{\nu}=0$, the conditions $\ka_{-1}= \ka_{-2} =  0$ are identical to the phase transitions in the MF approximation. Consequently, the FM phase is only defined in the regions where is the ground state, \ie, it is not a metastable phase nowhere at $T=0$. We plot the allowed regions of the FM phase, and the other phases, for the finite temperatures $T/T_0=0 , 0.05 , \dots , 0.3$ calculated numerically in Fig.~\ref{Adm.region}, which agrees with the previous discussion. The phase transitions of the FM phase does not depend on the temperature. On the other hand, its allowed region increases with respect to the temperature which would be the $(c_1 , c_2)$-values where the FM phase could be metastable. 
\subsubsection{P phase}
The $P$ phase is equal to the $\ket{2,0}$-phase, which has the same symmetries as the FM phase plus the time-reversal symmetry. Hence, besides to have the same density matrix as the FM phase, $\rho^{nc}= \sum_{m=-2}^2 \La_m \ket{2,m} \bra{2,m}   \, ,$ the additional symmetry yields $ \La_2 = \La_{-2}$ and $ \La_1 = \La_{-1}$. The allowed region of P at finite temperatures (see Fig.~\ref{Adm.region}) is restricted by the conditions of $\kappa_2 = \kappa_{-2} = 0$ (upper bound) and $\kappa_1 = \kappa_{-1} = 0$ (left-lower bound) where
\begin{align}
\ka_2^{(P)} = \ka_{-2}^{(P)} = & -\frac{c_2}{5}N + \mathcal{O}(\La_{\nu}) \, ,
\\
\ka_1^{(P)} = \ka_{-1}^{(P)} = & \left( 3c_1 - \frac{c_2}{5} \right)N + \mathcal{O}(\La_{\nu}) \, .
\end{align}
The left-lower bound does not coincide to the $P-S$ transition phase even at all temperatures, implying that the $P$ phase is metastable over the ground-phase region of the $S$ phase even at $T=0$.
\subsubsection{S phase}
The allowed region of the $S$ phase coincides to the zone where the nematic family is the ground phase at $T=0$. The left-lower and upper boundaries are given by $\ka_3^{(S)}= \ka_4^{(S)} = 0$, respectively, with
\begin{align}
\ka_3^{(S)} = & -\frac{c_2}{5}N + \mathcal{O}(\La_{\nu}) \, , 
\\
\ka_4^{(S)} = & \left(4c_1 -\frac{c_2}{5} \right) N + \mathcal{O}(\La_{\nu}) \, ,
\end{align}
with corresponding spin states given by the Eqs.~\eqref{eig.Sq1}-\eqref{eig.Sq2}
\begin{equation}
\bm{v}_{3} = (0,0,1,0,0)^{\text{T}} \, , \quad
\bm{v}_4 = (-1,0,0,0,1)^{\text{T}} / \sqrt{2} \, .
\end{equation}
\subsubsection{C case}
We plot in Fig.~\ref{Adm.region} the allowed regions at finite temperatures of the C phase. The left and lower boundaries are associated to the $\ka_1^{(C)},  \ka_3^{(C)} = 0$ conditions, respectively, where
\begin{align}
\ka_1^{(C)} = & \ka_2^{(C)} = \ka_5^{(C)}  = 2c_1 N + \mathcal{O}(\La_{\nu})
\, ,
\nonumber
\\
\ka_3^{(C)} 
= & \frac{2c_2}{5} N + \mathcal{O}(\La_{\nu})
\end{align}

The numerical calculations show the predicted triple-degeneracy of $\rho^{nc}$ over its allowed region.  Moreover, the spectrum of $\rho^{nc}$ is fourth-degenerated over the line $c_2 = 5c_1$. This result can be explained by the previous equations, where $\ka_1$ and $\ka_3$ are functionally equal when we substitute the conditions $c_2 = 5c_1$ and $\La_1 = \La_3$.
\section{Conclusions}
\label{Sec.Conc}
We have studied the emergent physics in the spin phases of BECs with a point group symmetry at finite temperatures. By taking advantage that the HF theory preserves the symmetries self-consistent, we use 
\rema{a method, based in the Majorana representation of mixed states,} to completely characterize the noncondensate fraction of atoms with the minimum degrees of freedom required for each point group symmetry. \remaRe{These characterizations, done in Sec.~\ref{Sec.Sphases}, can be applied to any model with self-consistent symmetries. In particular, we use them to } review the multipolar magnetic moments of two families of spin phases, called $\ket{f,m}$ and NOON, and of the exotic phases associated with the platonic solids. For this, some results about the anticoherence of spin states, quantity defined in quantum information theory, were useful. At last, we explored the allowed regions of each phase of the spin-2 BEC, which increase as the temperature increases. The implementation of this method is quite general and can be applied for any spin phase with a point group symmetry to reduce some degrees of freedom. Moreover, one could apply the method in others cases where the spatial part is no longer factorized, for example with dipolar spinor BEC~\cite{Lahaye_2009} or BEC with synthetic spin-orbit coupling~\cite{lin2011spin}, or even in time-dependent variations~\cite{PhysRevLett.77.5320}. 
\begin{acknowledgments}
E.S.-E. acknowledges support from the postdoctoral fellowships offered by DGAPA-UNAM and the IPD-STEMA program of the University of Liège. F.M. acknowledges the support of DGAPA-UNAM through the project PAPIIT No. IN113920.
\end{acknowledgments}
\appendix
\section{Characterization of the global phase factors in the Majorana representation for mixed states}
\label{App.phase}
The constellations plotted in the Fig.~\ref{MSC.Platonic} on the sphere of different radii characterize the density matrix of the noncondensate fraction up to the global phase factor of the $\bm{\rho}_{\si}$ vectors. However, one can associate the global phase factor of $\bm{\rho}_{\si}$ to an equivalence class of the set of the half of the points in the constellation. The whole details about this characterization is given in Ref.~\cite{Ser.Bra:20}. Here, we explain the basic notions and the schematic procedure.

Let us consider a vector $\bm{\rho}_{\si}$ of the Majorana representation of a mixed state $\rho$, which we write as a ket $\ket{\bm{\rho}_{\si}}$ to simplify the discussion. The state $\ket{\bm{\rho}_{\si}}$ has a Majorana constellation $\con{\si}$ with $2\si$ stars denoted by the tuple of unit vectors $(v_1 , \dots , v_{2\si} )$. The constellation $\con{\rho_\si}$ has antipodal symmetry, which implies that there exists subconstellations of $\si$ stars $\bm{c}=(\bm{v}_{\alpha_1} , \dots , \bm{v}_{\alpha_{\si}})$ such that 
\begin{equation}
\label{cond.cons}
\{ \bm{c} \} \cup \{ -\bm{c} \} = \con{\rho_\si} 
\, ,
\end{equation}
with $-\bm{c}=(-\bm{v}_{\al_1} , \dots , -\bm{v}_{\al_{\si}})$. In general, $\{ \bm{c} \}$ is not unique, and the other choices can be written with respect to $\bm{c}$ inverting the direction of some of its stars $\bm{\gamma} \bm{c} \equiv (\gamma_1 \bm{v}_1 , \dots , \gamma_{\si} \bm{v}_{\si})$, with $\gamma_{\al}=1$ or $-1$. Now, we can define a spin-$\si/2$ state $\ket{z_c}$ for each $\bm{c}$ via the Majorana representation. Analogously, the antipodal tuple $-\bm{c}$ has associated a spin-$\si/2$ state that we denote by $\ket{z_c^{\TRS}}\equiv \TRS \ket{z_c}$ and it is completely defined by the components of $\ket{z}$
\begin{equation}
\label{T.state}
\TRS \ket{z_c} = \sum_{m} (-1)^{f+m} t_{-m}^* \ket{f,m}
\, , \quad \text{for } \ket{z_c}=\sum_m t_m \ket{f,m} \, .
\end{equation} 
Both states define a $\si$-spin state, denoted by $\ket{Z_{c}}$, by its tensor product projected on the totally symmetric subspace 
\begin{equation}
\label{Eq.forc}
\ket{Z_c}= A \mathcal{P}_{\si} \left( \ket{z_c} \otimes \ket{z^{\TRS}_c} \right) \, , \quad \text{with } \mathcal{P}_{\si} = \sum_{m} \ket{\si,m} \bra{\si,m} \, ,
\end{equation}
where $A$ is a positive normalization factor. Surprisingly, $\ket{Z_c}$ is independent of the global phase factor of $\ket{z_c}$ \cite{Ser.Bra:20}. Moreover, $\ket{Z_c}$ is equal to one of the possible options $\pm \ket{\bm{\rho}_{\si}}$. Hence, we can associate to each choice of $\pm \ket{\bm{\rho}_{\si}}$, all the tuples $\bm{c}$ of $\si$ stars that defines the same spin state $\pm \ket{\bm{\rho}_{\si}}$ with the latter procedure. It turns out that the tuples $\bm{c}$ that define $\ket{Z_c} = \ket{\bm{\rho}_{\si}}$ differs among themselves only by an even number of stars \cite{Ser.Bra:20}. We then can define two equivalence classes between the subconstellations that satisfies Eq.~\eqref{cond.cons}
\begin{align}
& \left\{  \bm{\gamma} \bm{c} \subset \co^{(\si)} \Big|  \gamma_k= 1 \text{ or } -1 \, \text{ and }   \prod_{k=1}^{\si} \gamma_k =+ 1  \right\} \, ,
\nn
\\
& \left\{ \bm{\gamma} \bm{c} \subset \co^{(\si)} \Big|  \gamma_k= 1 \text{ or } -1 \, \text{ and }   \prod_{k=1}^{\si} \gamma_k =- 1   \right\} \, .
\label{class.cons}
\end{align}
Any element of any class produces the same spin state $\ket{\rho_{\si}}$ with Eq.~\eqref{Eq.forc} up to a global sign. On the other hand, only elements of the same class produce the same vector $\bm{\rho}_{\si}$, \ie, the same state and the same phase factor of $\bm{\rho}_{\si}$.
\begin{figure}[t!]
\begin{tabular}{|c|c|}
\hline
\hspace{1cm} $[\bm{c}] \, , \quad +\bm{\rho}_{\si}$ \hspace{1cm} & \hspace{1cm} $[-\bm{c}] \, , \quad -\bm{\rho}_{\si}$ \hspace{1cm}
\\
\hline
$\phantom{asdf}$ & $\phantom{asdf}$
\\[-0.3cm]
\includegraphics[scale=0.2]{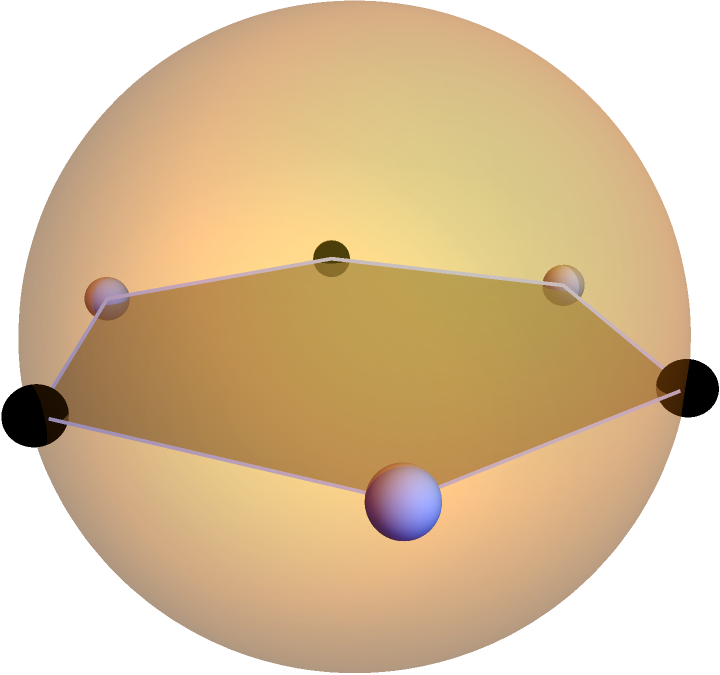}
&
\includegraphics[scale=0.2]{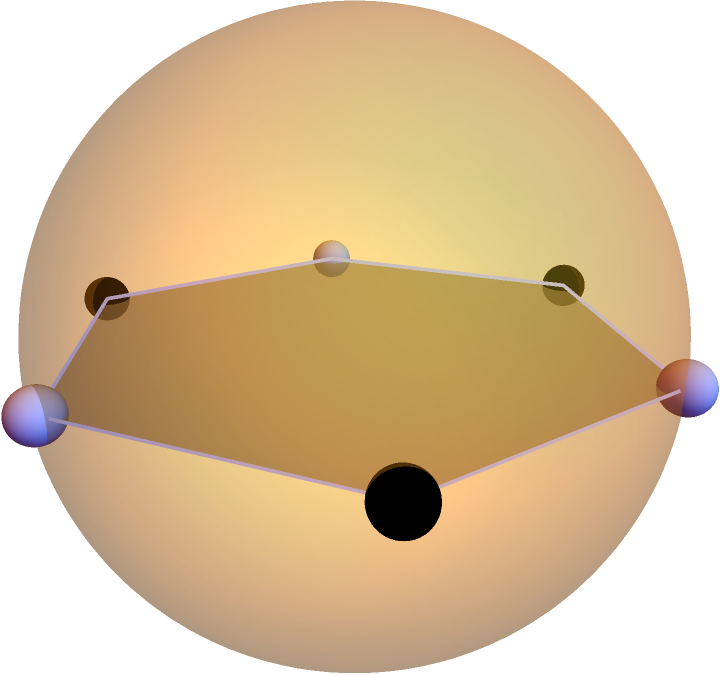}
\\[0.1cm]
\includegraphics[scale=0.2]{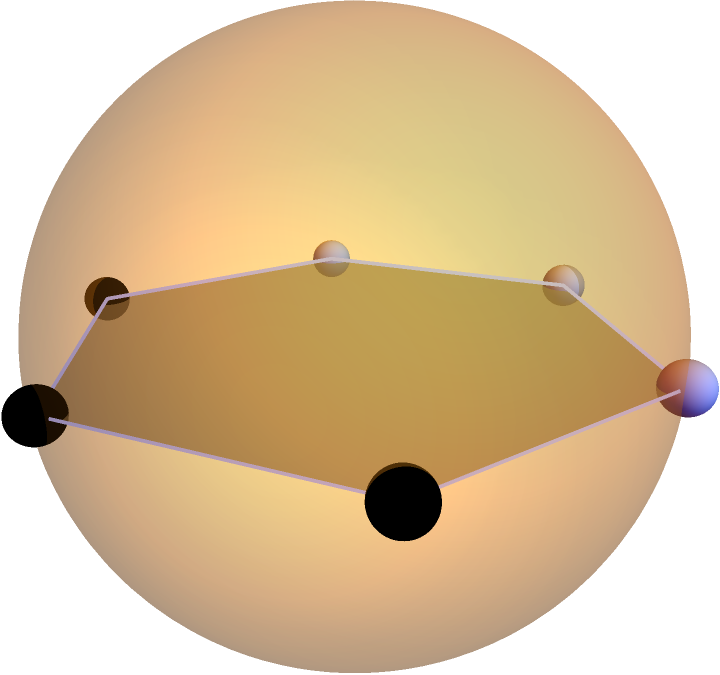}
&
\includegraphics[scale=0.2]{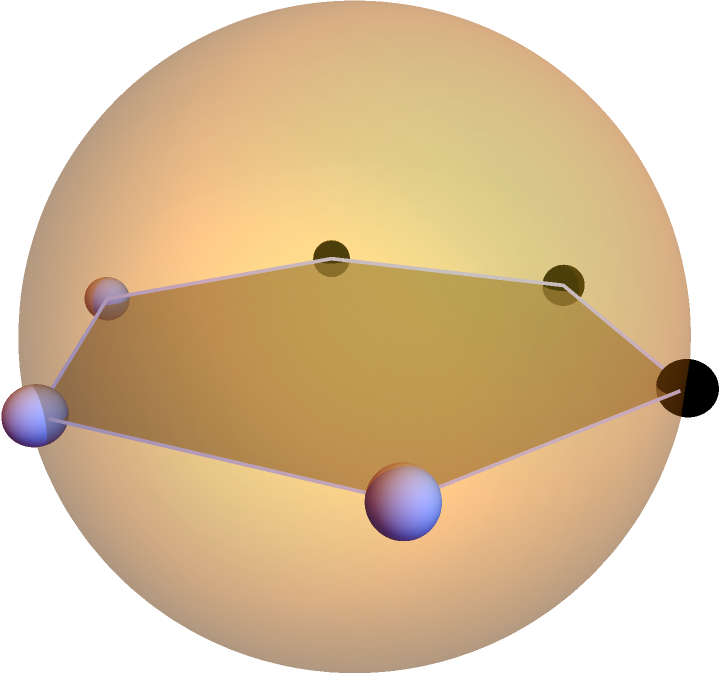}
\\[0.1cm]
\includegraphics[scale=0.2]{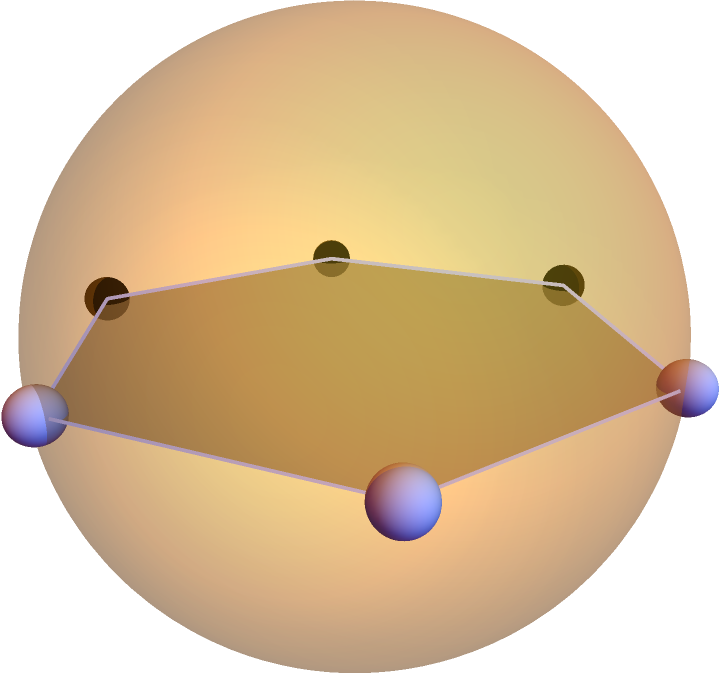}
&
\includegraphics[scale=0.2]{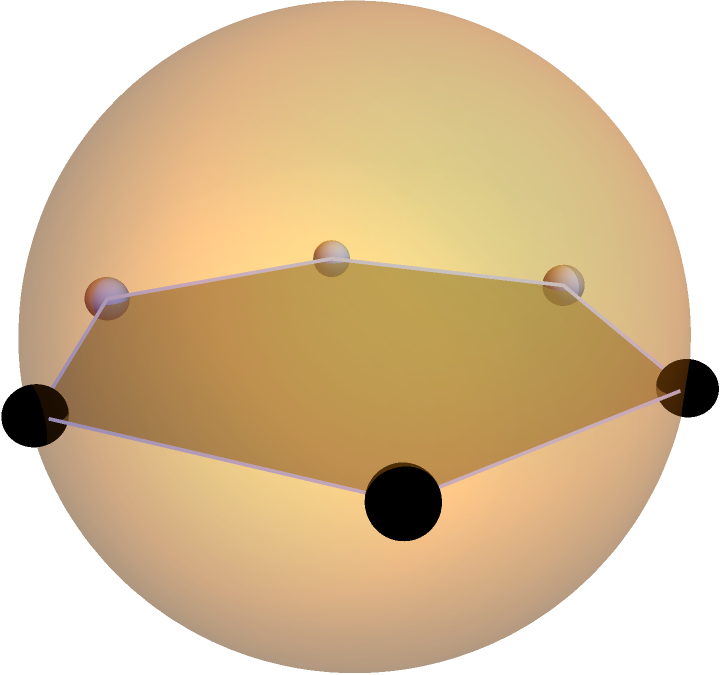}
\\[0.1cm]
\includegraphics[scale=0.2]{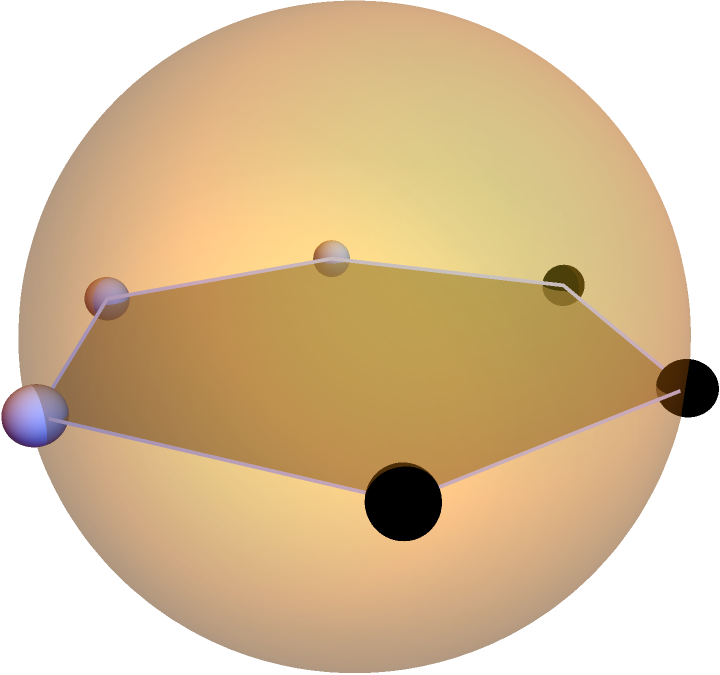}
&
\includegraphics[scale=0.2]{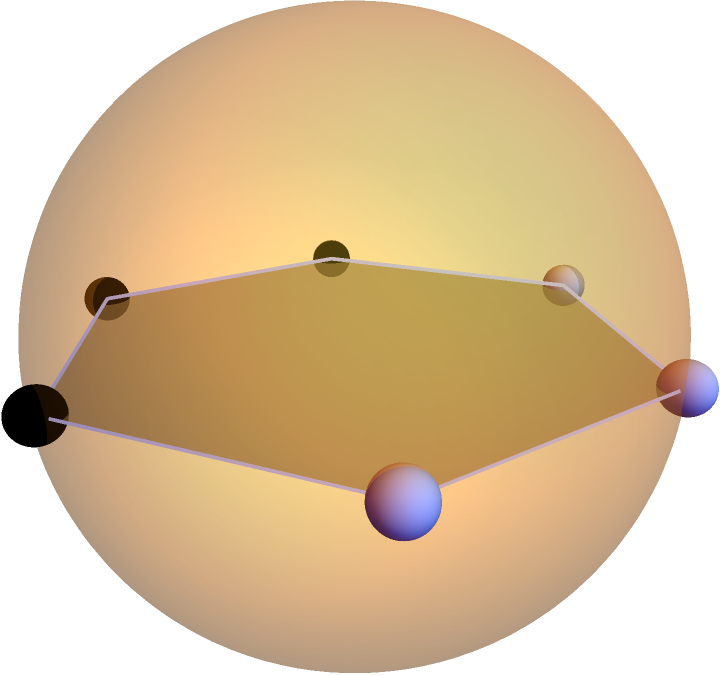}
\\
\hline
\end{tabular}
\caption{\label{Fig.Class} The two equivalence classes $[\pm \bm{c}]$ of the constellations associated to the Majorana stars of a hexagon. Each class is associated to the global sign of the vector $\pm \bm{\rho}_3$.}
\end{figure}

\def\sc2{2.7cm}
\begin{figure}[t!]
\large
\begin{tabular}{|c|}
\hline
$\mathcal{C}_{\rho^{nc}}$ 
\\
\hline
\begin{tabular}{c}
T ($f=2$) 
\\
$\si=3$ \hspace{2.2cm} $\si=4$
\\
\includegraphics[width=\sc2]{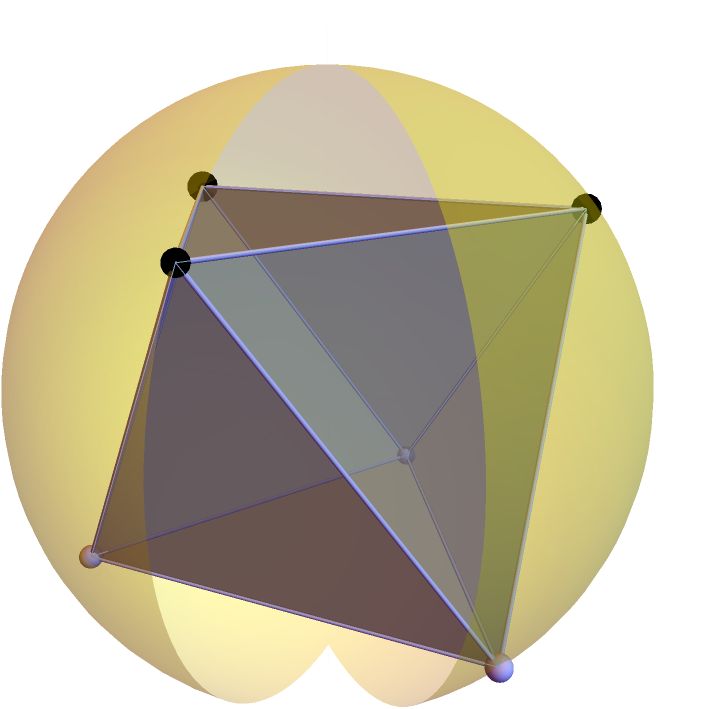}
\hspace{0.5cm}
\includegraphics[width=\sc2]{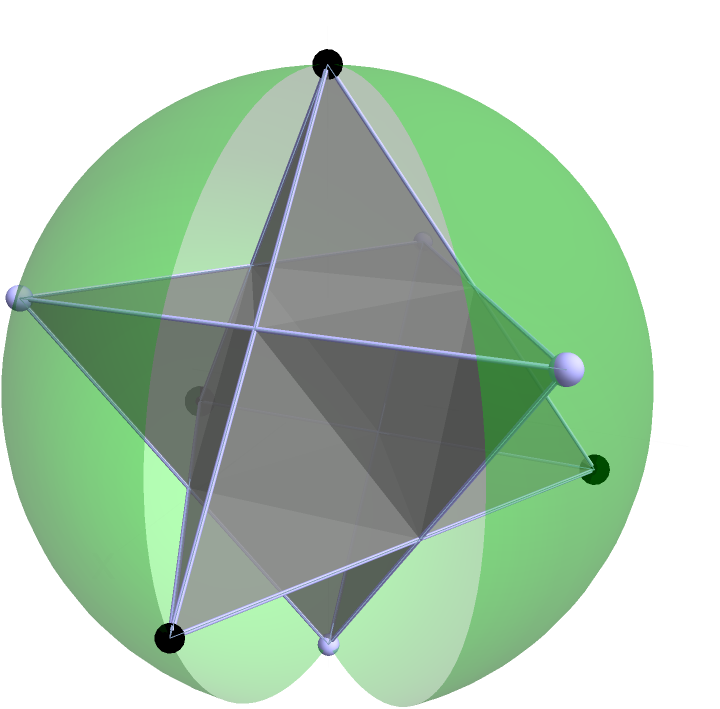}
\end{tabular}
\\
\hline
\begin{tabular}{c}
O ($f=3$) 
\\
$\si=4$ \hspace{2.2cm}  $\si=6$
\\
\includegraphics[width=\sc2]{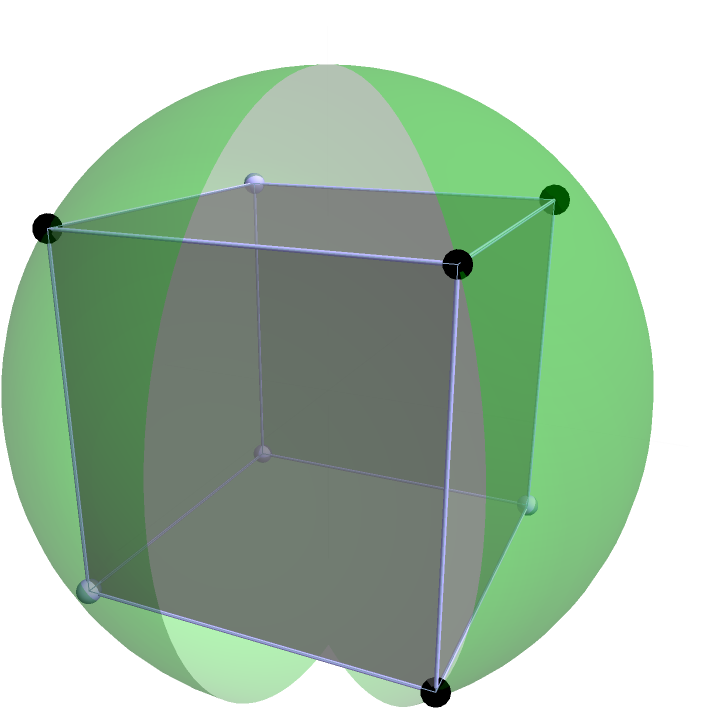} 
\hspace{0.5cm}
\includegraphics[width=\sc2]{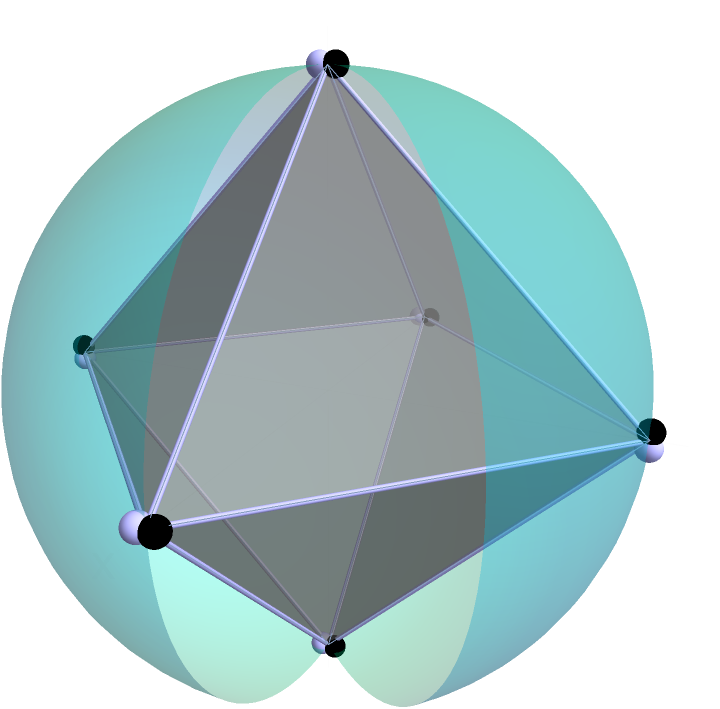}
\end{tabular}
\\
\hline
\begin{tabular}{c}
C ($f=4$)
\\
$\si=4$ \hspace{1.5cm} $\si=6$ \hspace{1.5cm} $\si=8$
\\
\includegraphics[width=\sc2]{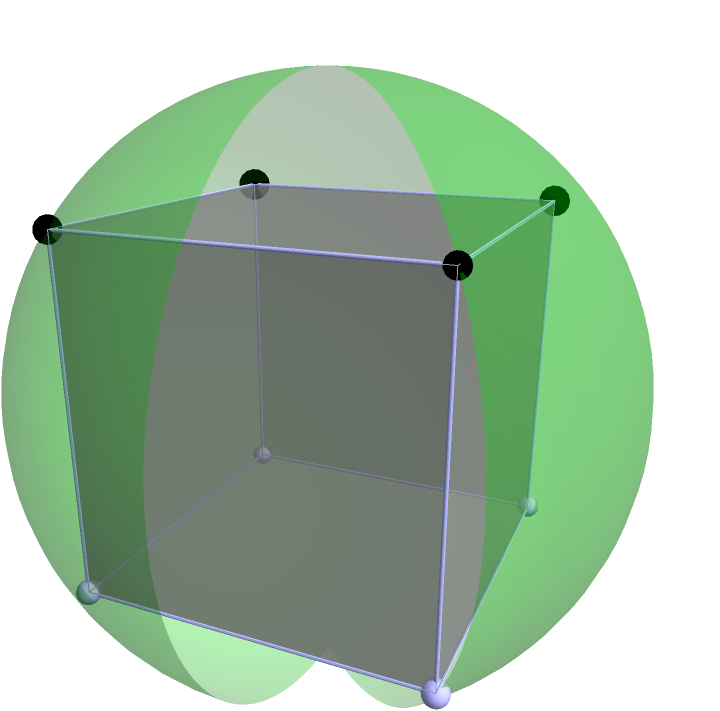}
\includegraphics[width=\sc2]{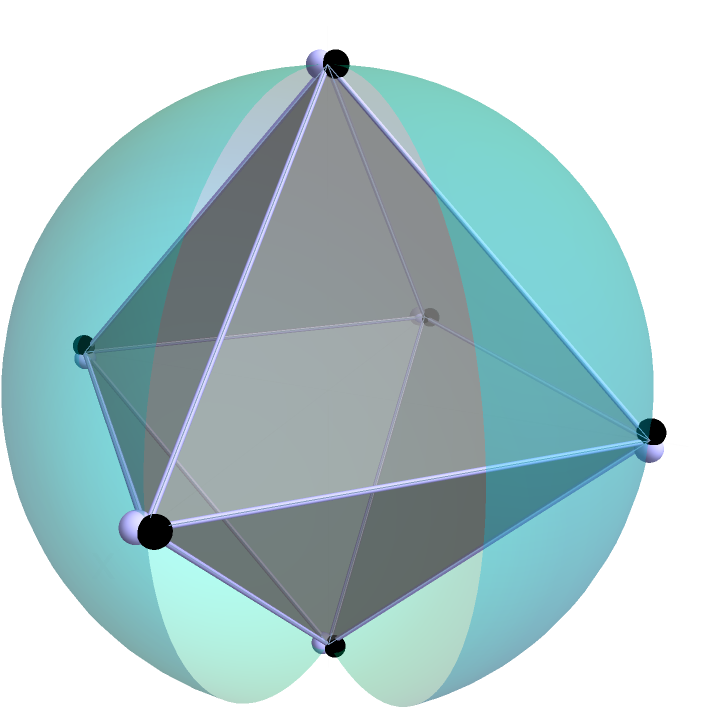}
\includegraphics[width=\sc2]{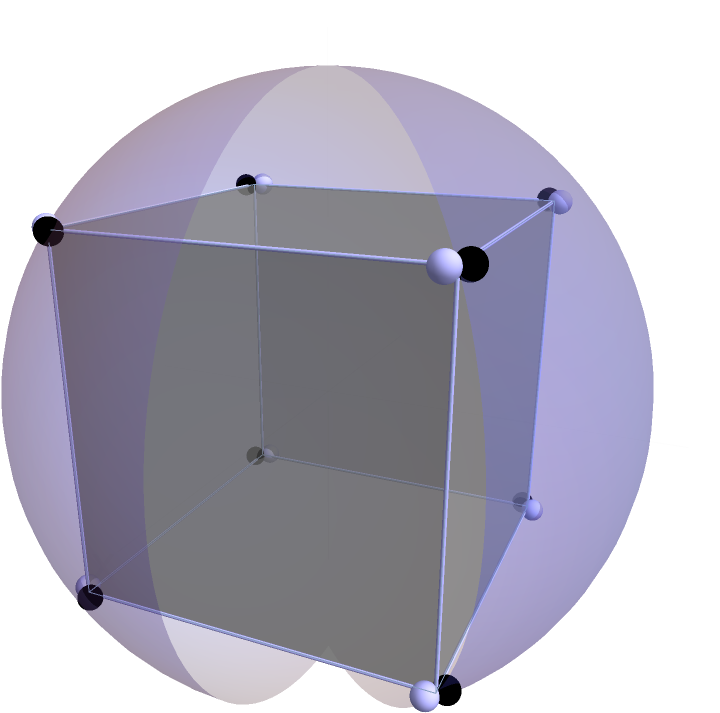}
\end{tabular}
\\
\hline
\begin{tabular}{c}
I ($f=6$) 
\\
$\si=6$ \hspace{1.5cm} $\si=10$ \hspace{1.5cm} $\si=12$
\\
\includegraphics[width=\sc2]{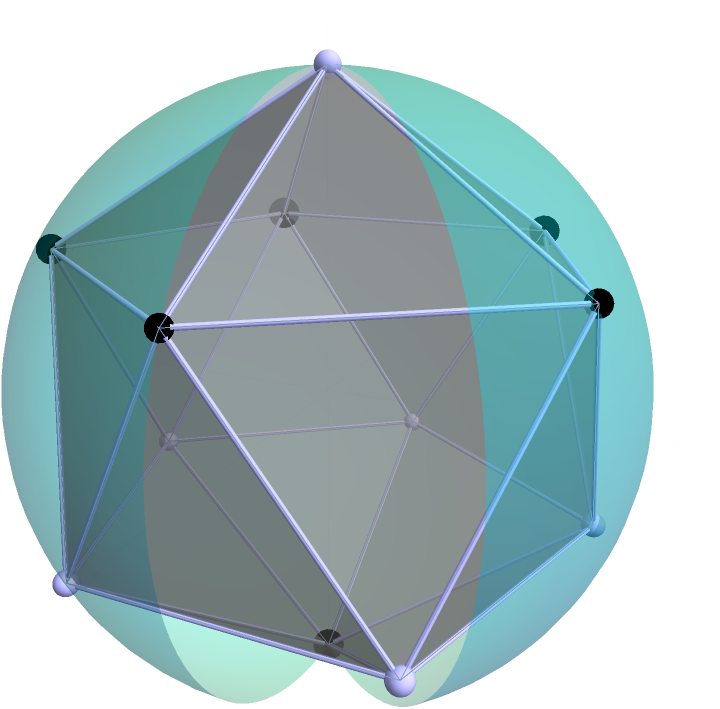}
\includegraphics[width=\sc2]{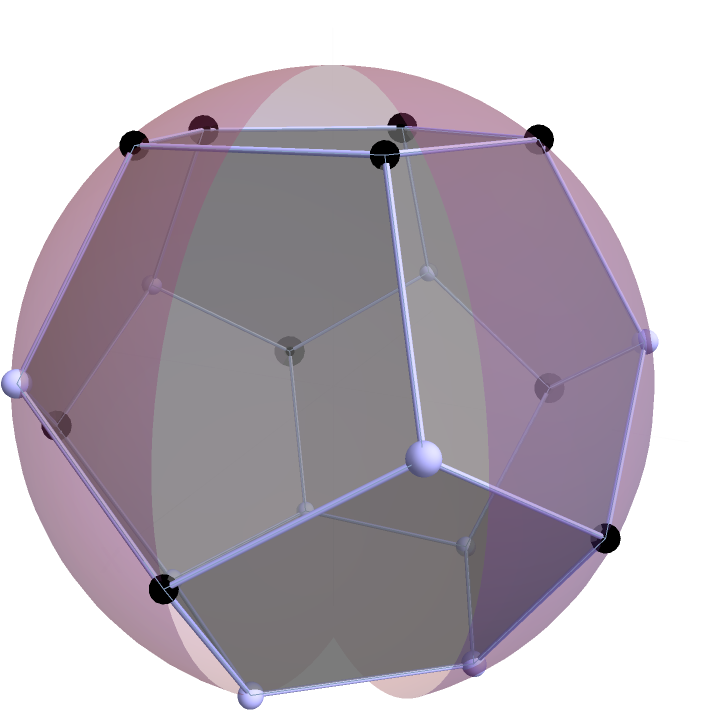}
\includegraphics[width=\sc2]{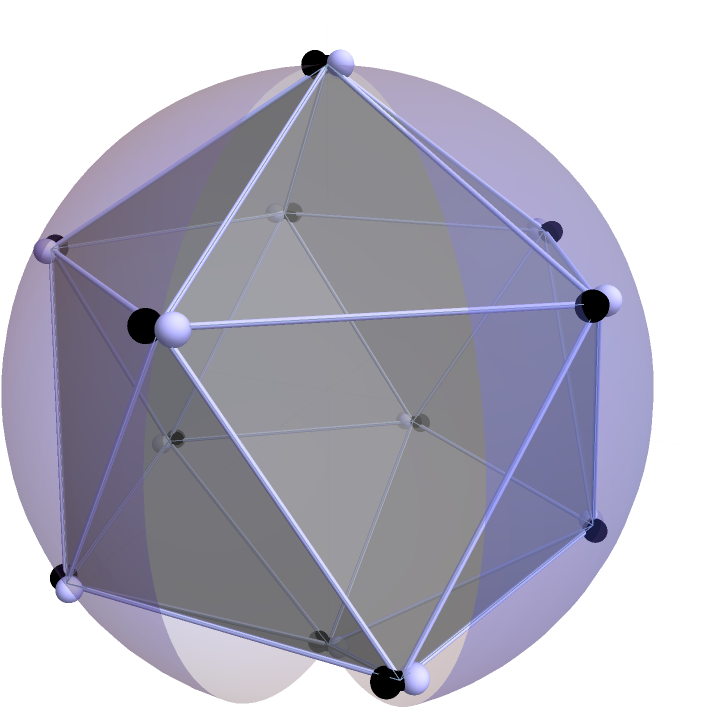}
\end{tabular}
\\
\hline
\end{tabular}
\caption{\label{Maj.Full} Majorana representation of the noncondensate fractions $\rho^{nc}$ associated to the platonic phases of the spinor BEC. The set of black stars constitutes a representative element of the equivalence class $[\bm{c}]$ of $\bm{\rho}_{\si}$.
} 
\end{figure}

To exemplify the method mentioned above, let us calculate the equivalence class associated to a vector $\bm{\rho}_3$ with 
\begin{equation}
\label{cons.hex}
\rho_{3m} = (\delta_{m,3} + \delta_{m, -3}) / \sqrt{2} \, .
\end{equation}
Its Majorana constellation consists of 6 points that conforms a regular hexagon on the Equator (see Fig.~\ref{Fig.Class}). Now, a tuple of stars $\bm{c}$ that satisfies Eq.~\eqref{cond.cons} consists of three points, let us say the black points of the first image of Fig.~\ref{Fig.Class}, which is associated to a spin-3/2 state equal to
\begin{equation}
\ket{z_c} = \frac{e^{i \beta}}{\sqrt{2}} \left( 
\left| \frac{3}{2} \, , \frac{3}{2} \right\rangle + \left| \frac{3}{2} \, , - \frac{3}{2} \right\rangle
\right) \, ,
\end{equation}
where $e^{i \beta}$ is a general global phase factor. The antipodal state is given by Eq.~\eqref{T.state}
\begin{equation}
\ket{z_c^{\TRS}} = \frac{e^{-i \beta}}{\sqrt{2}} \left( 
\left| \frac{3}{2} \, ,- \frac{3}{2} \right\rangle - \left| \frac{3}{2} \, ,\frac{3}{2} \right\rangle
\right) \, ,
\end{equation}
and the respective spin-3 state $\ket{Z_c}$ associated to the tuple $\bm{c}$ is calculated using Eq.~\eqref{Eq.forc}
\begin{align}
\ket{z_c}\otimes \ket{z_c^{\TRS}} =& \frac{1}{2} \left( \left| \frac{3}{2},-\frac{3}{2} \right\rangle \otimes \left| \frac{3}{2},-\frac{3}{2} \right\rangle  - \left| \frac{3}{2},\frac{3}{2} \right\rangle \otimes \left| \frac{3}{2},\frac{3}{2} \right\rangle
 \right. 
\nonumber
\\
+ & \left. \left| \frac{3}{2},\frac{3}{2} \right\rangle
\otimes \left| \frac{3}{2},-\frac{3}{2} \right\rangle
 -  \left| \frac{3}{2},\frac{3}{2} \right\rangle
\otimes \left| \frac{3}{2},-\frac{3}{2} \right\rangle
\right)
\nonumber
\\
\ket{Z_c} = & \frac{1}{\sqrt{2}} \left( \ket{3,3} + \ket{3,-3} \right) \, ,
\label{Cal.class}
\end{align}
where we use the basis transformation between coupled and decoupled angular momentum states of spin $3/2$. Therefore, the tuple $\bm{c}$ gives the same $\bm{\rho}_{3}$ given in Eq.~\eqref{cons.hex}, and then its equivalence class is given by all the constellations of three points that differs of  $\bm{c}$ by an even numbers of stars. All the possible cases are plotted in Fig.~\ref{Fig.Class}. We also include in the figure the other equivalence class $[-\bm{c}]$ which, by a similar calculation as in Eq.~\eqref{Cal.class}, gives the vector $-\bm{\rho}_3$.

In summary, a complete characterization of a mixed state is the collection of $2f$ constellations, each one on a sphere of radii $r_{\si}$ and with one of the two possible equivalence classes $[\pm \bm{c}]$ \eqref{class.cons}, with $\bm{c}$ a representative element. We plot in Fig.~\ref{Maj.Full} the examples of the complete characterization of the general mixed states with a symmetry associated to a platonic solid shown in Fig.~\ref{MSC.Platonic}. We use the vectors $\bm{\rho}_{\si}$ mentioned in the main text Eqs.~\eqref{Tet.cont},\eqref{Oct.cont}-\eqref{Ico.cont} and we assume that $r_{\si}\geq 0$. We represent a representative element of each equivalence class with the black stars.
\begin{widetext}
\section{Quadrupolar magnetic moment in terms of the tensor operators}
\label{App.mag}
In this appendix, we calculate the components $N_{\nu_1 \nu_2}$ by using the equations \eqref{ang.mom}-\eqref{prod.tens},
\begin{align}
N_{zz } = \cte^2 T_{10}^2
= \cte^2 \sum_{l,m} \chi(1,1,l,f) c^{lm}_{10,10} T_{1m}
=
\frac{1}{30} \Bigg(
10f(f+1) \sqrt{2f+1} T_{00} 
+  \sqrt{5 \frac{(2f+3)!}{(2f-2)!}} T_{20}
\Bigg)
\, ,
\end{align}
\begin{align}
N_{xx} = & \frac{\cte^2}{2} \left[ - \frac{2}{\sqrt{3}} \chi(1,1,0,f) T_{00} + 
\chi(1,1,2,f) \left( T_{22} + T_{2-2} - \sqrt{\frac{2}{3}} T_{20}  \right) 
\right]
\nonumber
\\
= & \frac{f(f+1)\sqrt{2f+1}}{3} T_{00} + 
\frac{1}{4\sqrt{30}}\sqrt{\frac{(2f+3)!}{(2f-2)!}}
 \left( T_{22} + T_{2-2} - \sqrt{\frac{2}{3}} T_{20}  \right) \, ,
\end{align}
\begin{align}
N_{yy} = &\frac{f(f+1)\sqrt{2f+1}}{3} T_{00} - 
\frac{1}{4\sqrt{30}}\sqrt{\frac{(2f+3)!}{(2f-2)!}}
 \left( T_{22} + T_{2-2} + \sqrt{\frac{2}{3}} T_{20}  \right) \, ,
\end{align}
\begin{align}
N_{xz}= N_{zx} = \frac{\cte^2}{2} \chi(1,1,2,f) \left( 
T_{2-1} - T_{21}
\right)
= \frac{4}{\sqrt{30}}\sqrt{\frac{(2f+3)!}{(2f-2)!}} \left( 
T_{2-1} - T_{21}
\right)
\, ,
\end{align}
\begin{align}
N_{yz} = N_{zy}= \frac{i \cte^2}{2} \chi(1,1,2,f) \left( 
T_{2-1} + T_{21}
\right)
= \frac{4i}{\sqrt{30}}\sqrt{\frac{(2f+3)!}{(2f-2)!}} \left( 
T_{2-1} + T_{21}
\right)
\, ,
\end{align}
\begin{align}
N_{xy} = N_{yx} = \frac{i \cte^2 \chi(1,1,2,f)}{2} \left( 
T_{2-2} - T_{22}
\right)
= \frac{i}{4\sqrt{30}}\sqrt{\frac{(2f+3)!}{(2f-2)!}} \left( 
T_{2-2} - T_{22}
\right) \, .
\end{align}
\section{Full expression of the $\kappa_{\nu}$ energies and the chemical potential of the phases of spin-2 BEC}
\label{spin2.kappa}
We enlist the chemical potential and the $\kappa_{\nu}$ energies following the same notation as in the main text.
\subsection{FM case}
\begin{align}
 \mu^{(FM)}  = & 
 \, c_0(N+\La_2)   
 +2c_1(2N+2\La_2-2\La_0-3\La_{-1}-4\La_{-2}) 
 + \frac{2c_2 \La_{-2}}{5} 
 \, .  
\nonumber
\\
\kappa_2^{(FM)}= & \, (c_0+4c_1)N^c \, ,
\nonumber
\\
\kappa_1^{(FM)}= & \, c_0 (\La_1 - \La_2) + c_1 ( -4\La_2 - 2\La_1 + 3\La_0 + \La_{-1} + 2\La_{-2}) 
+ \frac{2c_2}{5}(\La_{-1}- \La_{-2}) 
\, ,
\nonumber
\\
\kappa_0^{(FM)}= & \, c_0(\La_0-\La_2) +c_1 (-4N-4\La_2+3\La_1+4\La_0+9\La_{-1}+8\La_{-2})
+ \frac{2c_2}{5} (\La_0-\La_{-2})
\, ,
\nonumber
\\
\kappa_{-1}^{(FM)}= & \,
 c_0(\La_{-1}-\La_2) + c_1 (-6N -4\La_2 + \La_1+9\La_0+10\La_{-1}+14\La_{-2})
+ \frac{2c_2}{5} (\La_1-\La_{-2}) \, ,
\nonumber
\\
\kappa_{-2}^{(FM)}= & \, c_0(\La_{-2}-\La_2) + 2c_1 (-4N -2\La_2 + \La_1 + 4\La_0 + 7\La_{-1} + 10\La_{-2}) 
+ \frac{2c_2}{5}(N -\La_1-\La_0-\La_{-1}-2\La_{-2}) \, .
\end{align}
\subsection{P phase}
\begin{align}
\mu^{(P)} = & \,
c_0(N + \La_0) + 6 c_1 \La_1 + \frac{c_2}{5} \left(N + \La_0 - 2\La_1 - 2\La_2 \right)
\, ,
\nonumber
\\
\ka_2^{(P)} = & \,  \ka_{-2}^{(P)} = \, c_0( \La_2 -\La_0 ) + 4c_1 (\La_2 - \La_1) 
 - \frac{c_2}{5} \left(N +\La_0 - 2\La_1 -4\La_2 \right)
\, ,
\nonumber 
\\
\ka_1^{(P)} = & \, \ka_{-1}^{(P)} = \, c_0(\La_1 - \La_0) + c_1 \left( 3N - 11 \La_1 - 4 \La_2 \right)
  - \frac{c_2}{5} \left( N+ \La_0 - 4\La_1 - 2\La_2 \right) \, ,
\nonumber
\\
\ka_0^{(P)} = & \left( \frac{c_2+5c_0}{5} \right)N^c \, .
\end{align}
\subsection{S phase}
\begin{align}
\mu^{(S)} = & \, c_0 \left( N + \La_5 \right) + 2c_1 \left(\La_2 +2 \La_4 \right) + \frac{c_2}{5} \left(
N - 2\La_2 - \La_3 - \La_4 + \La_5
\right) 
\, ,
\nonumber 
\\
\kappa_1^{(S)} = & \, \kappa_2^{(S)} =
+c_0 \left( \La_2 - \La_5 \right)
+ c_1 \left( N -3 \La_2 +2 \La_3 -4 \La_4 \right)
+\frac{c_2}{5} \left( -N + 4 \La_2+ \La_3 + \La_4- \La_5 \right) \, ,
\nonumber 
\\
\kappa_3^{(S)} = & \,
c_0 \left( \La_3  - \La_5 \right) + 4 c_1 (\La_2 - \La_4) + \frac{c_2}{5} (-N + 2 \La_2 + 3 \La_3 + \La_4 - \La_5) \, ,
\nonumber
\\
\kappa_4^{(S)} = & \,
c_0 \left( \La_4 - \La_5 \right) - 4 c_1 (-N + 2 \La_2 + \La_3 + 2 \La_4) + 
 \frac{c_2}{5} (-N + 2 \La_2 + \La_3 + 3 \La_4 - \La_5) \, , 
 \nonumber
 \\
\kappa_5^{(S)} = & \, \frac{(5c_0 + c_2)}{5} N^c \, .
\end{align}
\subsection{C case}
The chemical potential is equal to 
\begin{align}
\mu^{(C)} = 
 c_0(N+ \La_4) + 6c_1 \La_1 + \frac{2c_2}{5} \La_3
 \, .
\end{align}
After we write $\rho^{nc}$ in terms of the $\La_{\nu}$, we can now calculate the HF equations given by the eigenvalues and eigenvectors equal to Eq.~\eqref{HF.nc},
\begin{align}
\ka_1^{(C)} = & \ka_2^{(C)} = \ka_5^{(C)}  =  c_0 (\La_1 - \La_4) + 2c_1 (N - 5\La_1) + \frac{2c_2}{5}(\La_1 - \La_3) 
\, ,
\nonumber
\\
\ka_3^{(C)} = & c_0 (\La_3 - \La_4) + \frac{2c_2}{5} (N - 3\La_1 - 2\La_3) \, , 
\nonumber
\\ 
\ka_4^{(C)}= & c_0 N^c \, .
\label{Ener.C}
\end{align}
\end{widetext}
\bibliographystyle{apsrev4-2}
\bibliography{refs_joint_long.bib}
%
%
%
%
%
%
\end{document}